\documentclass[conference, compsoc]{IEEEtran}

\usepackage{float}
\usepackage{tikz}
\usetikzlibrary{arrows.meta, positioning, calc, decorations.pathreplacing, tikzmark}
\usepackage{comment, graphicx, booktabs, algorithm, algpseudocode, xspace, listings, subcaption, multirow, siunitx, amsmath, amsthm, xtab}
\usepackage[inline]{enumitem}
\usepackage{array}
\usepackage{threeparttable}
\usepackage{soul}
\usepackage{pbox}
\usepackage{balance}
\usepackage[many]{tcolorbox}
\newtcolorbox[auto counter]{mybox}[2][]{enhanced jigsaw, breakable, #1}

\definecolor{junkbyte}{RGB}{255,200,200}  
\definecolor{patterncolor}{RGB}{255,255,150} 
\definecolor{skipcolor}{RGB}{144,238,144} 

\usepackage{xcolor}
\usepackage{amssymb}
\usepackage{pifont}
\usepackage{seqsplit}
\usepackage{flushend}

\usepackage{tabularray}
\UseTblrLibrary{booktabs}

\PassOptionsToPackage{hyphens}{url}
\usepackage{xurl}
\usepackage[hidelinks]{hyperref}     
\usepackage[hyphens]{url}
\usepackage{longtable}
\usepackage{booktabs}
\usepackage{tabularx}
\usepackage{array}
\algnewcommand\algorithmicswitch{\textbf{switch}}
\algnewcommand\algorithmiccase{\textbf{case}}
\algnewcommand\algorithmicassert{\texttt{assert}}
\algnewcommand\Assert[1]{\State \algorithmicassert(#1)}
\algdef{SE}[SWITCH]{Switch}{EndSwitch}[1]{\algorithmicswitch\ #1\ \algorithmicdo}{\algorithmicend\ \algorithmicswitch}
\algdef{SE}[CASE]{Case}{EndCase}[1]{\algorithmiccase\ #1}{\algorithmicend\ \algorithmiccase}
\algtext*{EndSwitch}
\algtext*{EndCase}

\graphicspath{{figure/}}

\newcommand{\tool}{\textsc{ABLE}\xspace}

\lstdefinelanguage{PromptFuzz}{
  sensitive = true,
  keywords={call, load, update, assert, file},
  morekeywords=[2]={mut*, const*},
  comment=[l]{//},
  morestring=[b]',
  morestring=[b]"
}

\begin{document}

\date{}

\title{A Large Language Model Approach to Generating Bypass Rules for Malware Evasion in Analysis Sandbox}

\author{
{\rm Zhiyong Sui, Lamine Noureddine, Mst Eshita Khatun,}\\
{\rm Sideeq Bello, Justin Woodring, Aisha Ali-Gombe}\\
Louisiana State University\\
\{zsui1, lnoureddine, mkhatu3, sbell49, jwoodr7, aaligombe\}@lsu.edu
}

\maketitle

\begin{abstract}
Sandbox evasion remains a critical challenge for automated malware analysis, as modern malware employs environment checks to detect analysis platforms and suppress malicious behavior. Existing approaches rely on manually crafted bypass rules that require deep reverse engineering of each evasion mechanism—an approach that cannot scale against rapidly evolving evasion techniques. In this paper, we leverage large language models (LLMs) to automatically generate YARA rules that bypass evasion checks in sandbox environments. We propose ABLE, which analyzes execution traces from malware terminated due to potentially evasive behavior and employs multiple reasoning strategies to generate targeted bypass rules. To address syntactic errors and improve the efficacy of the bypass rules in the LLM outputs, we introduce an auto-sanitization pipeline and feedback-driven iterative refinement. We evaluate ABLE on 334 real-world malware samples across four open-weight LLMs. ABLE achieves a 79\% bypass success rate, with iterative refinement contributing 29.5\% of successful cases. Compared to existing analysis platforms, ABLE identifies 47\% more malware family classifications and exposes previously hidden behaviors.
\end{abstract}

\section{Introduction}
\label{sec:intro}
The proliferation of stealthy malware employing advanced evasion techniques continues to increase in both scale and complexity. Recent studies have shown that modern malware are engineered to conceal their behavior, leveraging various evasion tactics such as code obfuscation and anti-analysis to condition their behavior on environmental signals, selectively withholding logic bombs or infection payloads until specific criteria are met \cite{vmray-evasion,gorter2023enviral,kruegel2015evasive,bulazel2017survey}. To address this challenge, sandbox-based analysis has emerged as a primary mechanism, supporting both human-in-the-loop investigation and fully automated execution environments. Yet despite their widespread adoption, existing sandboxes lack effective mechanisms to counter evasion techniques and thus often fail to reliably elicit and observe the full range of malicious behaviors. Empirical studies \cite{kruegel2015evasive,phillipslongitudinal,sui6295224obfuscation} show that a substantial fraction of malware executed in sandbox environments actively employs evasion strategies that significantly undermine analysis fidelity. These failures are not incidental; modern malware is explicitly designed to detect sandbox artifacts and suppress or alter its behavior accordingly, exposing systemic weaknesses in conventional sandbox-based analysis approaches. Particularly, recent and sophisticated malware families such as Saint Bear\cite{unit42-outsteel-saintbot}, Bazar \cite{cybereason2020bazar}, Egregor \cite{nhsdigital2020egregor}, Hancitor \cite{anubhav2016hancitor}, Metamorfo\cite{porolli2021spalax}, StealC malware \cite{sekoia2023stealc}, whose primary function is similar to Raccoon \cite{esentire2022raccoon}, and Mars stealers \cite{osipov2022mars} and StoneDrill\cite{kaspersky2017shamoon} incorporate complex custom anti-analysis and evasion mechanisms that probe execution environments for hardware artifacts, operating system features, and runtime configurations before activating malicious logic. These checks are often distributed across multiple stages of the malware lifecycle, allowing behavior to be delayed, conditionally triggered, or entirely concealed\cite{barone2022reverse}. As a result, malware may appear benign or inactive during sandbox execution despite being fully functional in its intended deployment environment. 

In practice, as shown in the 2021 study by Wong et al. \cite{yong2021inside}, expert analysts in advanced tiers employ debugging and binary patching techniques to bypass or remove evasion logic and force malware to execute its malicious behavior. Common practices include using debuggers to skip or manipulate evasion checkpoints such as anti-virtualization or environment-check code paths, often with the aid of signatures such as YARA rules. The goal is for the encoded rules to trigger specific alerts or actions during execution. These rules are typically derived through extensive manual reverse engineering, inspection of dynamic execution traces, static binary, or in-memory code segments. However, this approach is fundamentally limited. First, modern malware often embeds multiple, layered evasion mechanisms that are distributed across different execution stages, making it impractical to identify and encode all relevant triggers using static or manually crafted YARA rules. Second, the reliance on expert-driven analysis to discover evasion logic and create YARA rules does not scale to the volume, diversity, and rapid evolution of contemporary malware. Finally, manual rule creation is inherently error-prone: incomplete or overly specific rules can miss evasive behaviors, while overly broad rules may introduce false positives or unintended execution paths. Therefore, while this manual approach remains state-of-the-art and enables analysis progression in practice, the study by Wong et al. also shows that executing all evasive stages and checkpoints of modern malware, and reliably exposing evasive behavior at scale, remains highly challenging, highlighting fundamental limitations of the method~\cite{yong2021inside}.

Therefore, we introduce \tool, an automated feedback-driven framework that leverages LLM semantic reasoning strategies to explore  hidden malicious behaviors in malware by analyzing dynamic execution traces. Unlike existing approaches, \tool detects evasive checkpoints, automatically generates YARA rules, and iteratively intervenes during execution with control flow manipulation actions, eliminating the need for extensive manual debugging or patching. The workflow of \tool consists of (1) an LLM-guided YARA rule generation, (2) a rule validation and repair sanitizer, and (3) an iterative sandbox execution and behavioral validator feedback engine. Through this proposed automated anti-analysis bypass, \tool can systematically find previously hidden behaviors by identifying and bypassingmultiple evasive checkpoints, thus achieving broader behavioral coverage than existing approaches. Unlike existing LLM-assisted tools for YARA rule generation~\cite{zhang2025automatically,patsakis2024assessing,khalifa2024sama,wang2025llmdyara}, \tool is designed to improve sandbox analysis fidelity by enabling the extraction of critical malware indicators and evasion tactics.

The evaluation of \tool demonstrates both the effectiveness of the end-to-end pipeline for real-world malware analysis and the contribution of its individual components. Experimental results on 334 real-world malware across 13,778 sandbox executions show that \tool achieves a 79\% bypass success rate, with iterative rule refinement accounting for 29.5\% of successful cases. Compared to existing analysis platforms, \tool outperforms prior approaches by identifying 47\% more malware family classification and revealing previously hidden malicious behaviors. 

In summary, we make the following salient contributions:
\begin{enumerate}
    \item An open-source automated framework that bypasses malware anti-analysis via iterative, feedback-driven sandbox interventions, integrating LLM-assisted YARA rule generation and actions, and compatible with existing sandbox and debugger-based engines such as CAPE.
    \item Expose hidden behaviors and signatures for known malware families through iterative anti-analysis bypass, which were not observable under conventional sandbox execution.
   \item Demonstrate that LLMs can effectively assist in overcoming challenges even for experienced malware analysts, as previously documented by Wong et al. \cite{yong2021inside}, by automating evasion identification and bypass workflows.
\end{enumerate}
\section{Background and Motivation}
\label{sec:background}
\subsection{Sandbox-based Analysis}
Traditionally, dynamic analysis techniques execute malware in controlled environments to observe runtime behavior, log API calls, network communications, file operations, and registry modifications \cite{egele2012survey}. Over time, sandbox-based analysis has been developed to simplify dynamic analysis by providing controlled environments that mimic real victim setups. These sandboxes have evolved from simple containers into fully automated execution environments, plugged with features such as GUI interaction, event injection, networking capabilities, and debugging tools, which significantly enhance their analytical effectiveness. Popular analysis sandbox include CAPE \cite{cape2024}, Cuckoo \cite{guarnieri2013cuckoo}, ANY.RUN \cite{anyrun2024}, and commercial solutions like Joe Sandbox \cite{joesandbox2024} and VMRay \cite{vmray2024}. Our approach takes a different direction: rather than improving post-sandbox reverse engineering, we enhance the sandbox's analytical capability itself by leveraging LLMs to automatically generate bypass rules—without requiring any modification to the underlying sandbox infrastructure.

However, malware authors actively develop techniques to detect sandbox environments and suppress malicious behavior\cite{galloro2022systematical,miramirkhani2017spotless,phillips2022sterilized,phillipsspvexec}. Such evasion mechanisms include:
(1) hardware fingerprinting via WMI queries that return empty results in VMs (e.g., Win32\_Fan, CIM\_TemperatureSensor), (2) timing checks using RDTSC to detect execution delays, (3) user activity detection requiring mouse movements or keyboard input, and (4) environment checks for known VM artifacts or sandbox usernames \cite{afianian2019malware}. Our evaluation of four major sandboxes (CAPE, AnyRun, Hybrid, QiAnXin) against al-khaser \cite{alkhaser}, a public evasion toolkit, shows that 17 of 228 checks evade at least one platform (Appendix~\ref{appendix:alkhaser}). 

\subsection{YARA Rules \& Debugger Actions}
YARA is a pattern-matching specification designed to identify text strings, binary or hex streams, and regular expressions in files, code, debug traces, or memory images \cite{raff2020automatic}. It is a standard tool used in malware analysis to identify and classify malicious software by describing its low-level characteristics and Indicators of Compromise (IoCs). By design, YARA is used to set up alerts on known evasion checkpoints determined through manual analysis. However, automated sandboxes, such as the CAPE sandbox, have extended traditional YARA rule matching to incorporate debugger actions that can modify program execution when the specified condition is met \cite{cape2024}. 

When a YARA rule matches a binary pattern during execution, an integrated sandbox debugger, e.g., CAPEMON\cite{capemon}, can perform actions such as \texttt{skip} (advance instruction pointer past the matched bytes), \texttt{wret} (force function return), or \texttt{setcf} (set CPU flags).
This mechanism enables analysts to bypass runtime evasion checkpoints. A typical bypass rule specifies: (1) a hex pattern identifying the evasion code location, (2) an offset from the pattern start to the target instruction, and (3) the action to perform. However, crafting effective rules requires understanding the exact binary layout, evasion logic, and appropriate bypass points — a manual process that does not scale.

\begin{figure}[!t]
\centering
\begin{lstlisting}[language=C, basicstyle=\scriptsize\ttfamily, numbers=left, numberstyle=\tiny, frame=single, breaklines=true]
void main_function() {
    // Stage 1: Initialization
    decrypt_string();
    resolve_WindowsAPI();
    string_convertion(&lpName, szAgent);
    // Stage 2: Anti-analysis checks
    checkCurrentProcess_isWritable();
    takeScreenshot();
    get_RAM_capacity(); // VM detection
    checkLangRU(); // Geofencing
    is_windows_defender_emulator();
    // Stage 3: Build mutex
    UserName_heap = GetUserName_heap();
    ComputerName = GetComputerName_heap();
    mutex_name = str_concat("HAL9TH", ComputerName, UserName_heap);
    // Stage 4: Single-instance check
    while (1) {
        hEvent = OpenEventA(EVENT_ALL_ACCESS,FALSE, mutex_name);
        if (!hEvent) break;
        CloseHandle(hEvent);
        Sleep(6000);
    }
    EventA = CreateEventA(NULL, FALSE, FALSE, mutex_name);
    // Stage 5: License check
    check_LicenseExpirationTime();
    // Stage 6: C2 payload
    MainFunctionC2_interaction();
    CloseHandle(EventA);
    ExitProcess(0);
}
\end{lstlisting}
\centerline{\footnotesize (a) Decompiled main function of StealC.}

\begin{lstlisting}[language=C, basicstyle=\scriptsize\ttfamily, numbers=left, numberstyle=\tiny, frame=single, breaklines=true]
void check_LicenseExpiration() {
    SYSTEMTIME current, exp;
    FILETIME cur_ft, exp_ft;
    GetSystemTime(&current);
    // Hardcoded expiration
    char* date = decrypt_date();
    sscanf(date, "%hu/%hu/%hu", &exp.wMonth, &exp.wDay,  &exp.wYear);
    // 04/02/2023
    SystemTimeToFileTime(&current, &cur_ft);
    SystemTimeToFileTime(&exp, &exp_ft);
    // Kill switch
    if (CompareFileTime(&cur_ft, &exp_ft) > 0) {
        ExitProcess(0);
    }
}
\end{lstlisting}
\centerline{\footnotesize (b) License expiration check.}

\definecolor{patterncolor}{RGB}{255,255,150}
\begin{lstlisting}[basicstyle=\scriptsize\ttfamily, numbers=left, numberstyle=\tiny, frame=single, escapeinside={(*@}{@*)}]
0040D077 (*@\colorbox{patterncolor}{53}@*)                     db 53h
0040D078 (*@\colorbox{patterncolor}{57 56 FF 15}@*)            dd 15FF5657h
0040D07C DC 35 61 00            dd offset dword_6135DC
0040D080 (*@\colorbox{patterncolor}{3B C7 75 E1 53 57 57 57}@*) dd 0E175C73Bh
0040D088 (*@\colorbox{patterncolor}{FF 15}@*)                  db 0FFh, 15h
0040D08A (*@\colorbox{patterncolor}{24 37 61 00}@*)             dd offset dword_613724
0040D08E (*@\colorbox{patterncolor}{8B F0}@*)                  dw 0F08Bh
0040D090 (*@\colorbox{patterncolor}{74 03 75 01}@*) (*@\colorbox{pink}{B8}@*) (*@\colorbox{lime}{E8 30 FE FF FF}@*) dd ...
\end{lstlisting}
\centerline{\footnotesize (c) IDA disassembly showing anti-disassembly artifacts.}
\caption{Static analysis of StealC malware.}
\label{fig:static_analysis}
\end{figure}

\begin{figure*}[t]
\centering
\definecolor{junkbyte}{RGB}{255,200,200}
\definecolor{patterncolor}{RGB}{255,255,150}
\definecolor{skipcolor}{RGB}{144,238,144}
\definecolor{bypassClight}{RGB}{180,220,220}
\definecolor{bypassAlight}{RGB}{255,200,150}
\definecolor{bypassBlight}{RGB}{144,238,144}
\definecolor{skipcolorlight}{RGB}{220,190,255}

\begin{minipage}[t]{0.4\textwidth}
\begin{lstlisting}[basicstyle=\scriptsize\ttfamily, numbers=left, numberstyle=\tiny, frame=single, escapeinside={(*@}{@*)}]
0017D084  (*@\colorbox{patterncolor}{53 57 57 57}@*)        push ebx; push edi x3
0017D088  (*@\colorbox{patterncolor}{FF 15}@*) 24 37 38 00  call [CreateEventA]
0017D08E  (*@\colorbox{patterncolor}{8B F0}@*)              mov esi, eax
0017D090  (*@\colorbox{patterncolor}{74 03}@*)              je 0017D095h
0017D092  (*@\colorbox{patterncolor}{75 01}@*)              jne 0017D095h
0017D095  (*@\tikzmark{callCECA}\colorbox{lime}{E8 30 FE FF FF}@*)    call 0017CECAh
0017D09A  (*@\colorbox{patterncolor}{74 03 75 01 B8}@*)     je; jne; ...
\end{lstlisting}%
\begin{tikzpicture}[remember picture, overlay]
    \draw[-{Stealth[length=2mm]}, blue, thick]
        ([xshift=5.0cm, yshift=0.08cm]pic cs:callCECA)
        node[anchor=west, font=\tiny\sffamily\bfseries, text=blue] {to (d)}
        -- ([xshift=4.0cm, yshift=0.08cm]pic cs:callCECA);
\end{tikzpicture}
\centering\footnotesize (c') Clean disassembly reveals call to 0017CECA.
\begin{lstlisting}[basicstyle=\scriptsize\ttfamily, numbers=left, numberstyle=\tiny, frame=single, escapeinside={(*@}{@*)}]
rule Bypass_Expiration_Check {
  meta:
    description = "Bypass time-based evasion"
    cape_options = "bp0=$anti+17,action0=skip,count=1"
  strings:
    $anti = { (*@\colorbox{patterncolor}{53 57 57 57 FF 15 [4] 8B F0 74 03 75 01}@*)
              (*@\colorbox{junkbyte}{B8}@*) (*@\colorbox{lime}{E8 [4]}@*)  (*@\colorbox{patterncolor}{74 03 75 01 B8}@*) }
  condition:
    uint16(0) == 0x5A4D and all of them
}
\end{lstlisting}
\centering\footnotesize (c'') YARA rule with bypass action.

\end{minipage}%
\hfill
\begin{minipage}[t]{0.58\textwidth}
\begin{lstlisting}[basicstyle=\scriptsize\ttfamily, numbers=left, numberstyle=\tiny, frame=single, escapeinside={(*@}{@*)}]
0017CECA  (*@\tikzmark{funcStart}\colorbox{bypassClight}{55}@*)                 push ebp
0017CECB  8B EC              mov ebp, esp
0017CECD  83 E4 F8           and esp, FFFFFFF8h
0017CED0  83 EC 44           sub esp, 44h
0017CED3  57	             push edi
0017CED4  33 C0	             xor eax, eax
          ...                ; initialize local variables
0017CF0D  50	             push eax
0017CF0E  FF 15 DC363800     call [GetSystemTime]
0017CF18  E8 DE FE FF FF     call 0017CDFBh     ; decrypt_date
0017CF34  FF 15 5C373800     call [sscanf]
0017CF50  FF 15 E4363800     call [SystemTimeToFileTime]
0017CF60  FF 15 E4363800     call [SystemTimeToFileTime]
0017CF6A  3B 44 24 14        cmp eax, [esp+14h] ; compare time
0017CF6E  (*@\tikzmark{jcInstr}\colorbox{bypassAlight}{72 14}@*)              jc 0017CF84h       ; not expired
0017CF70  (*@\tikzmark{jaInstr}\colorbox{bypassBlight}{77 0A}@*)              jnbe 0017CF7Ch     ; expired!
0017CF72  8B 44 24 08        mov eax, [esp+08h]
0017CF76  3B 44 24 10        cmp eax, [esp+10h]
0017CF7A  76 08              jbe 0017CF84h
0017CF7C  (*@\tikzmark{pushZero}\colorbox{skipcolorlight}{6A 00}@*)              push 0
0017CF7E  (*@\colorbox{skipcolorlight}{FF 15 BC363800}@*)  call [ExitProcess]
0017CF84  5F                 pop edi
0017CF85  8B E5              mov esp, ebp
0017CF87  5D                 pop ebp
0017CF88  C3                 ret
\end{lstlisting}%
\begin{tikzpicture}[remember picture, overlay]
    \draw[-{Stealth[length=2mm]}, teal, thick]
        ([xshift=5.8cm, yshift=0.08cm]pic cs:funcStart)
        node[anchor=west, font=\tiny\sffamily\bfseries, text=teal] {Wret}
        -- ([xshift=1.6cm, yshift=0.08cm]pic cs:funcStart);
    \draw[-{Stealth[length=2mm]}, orange, thick]
        ([xshift=5.8cm, yshift=0.08cm]pic cs:jcInstr)
        node[anchor=west, font=\tiny\sffamily\bfseries, text=orange] {SetCF/JMP}
        -- ([xshift=4.3cm, yshift=0.08cm]pic cs:jcInstr);
    \draw[-{Stealth[length=2mm]}, green!50!black, thick]
        ([xshift=5.8cm, yshift=0.08cm]pic cs:jaInstr)
        node[anchor=west, font=\tiny\sffamily\bfseries, text=green!50!black] {skip}
        -- ([xshift=4.2cm, yshift=0.08cm]pic cs:jaInstr);
    \draw[-{Stealth[length=2mm]}, violet, thick]
        ([xshift=5.8cm, yshift=0.08cm]pic cs:pushZero)
        node[anchor=west, font=\tiny\sffamily\bfseries, text=violet] {skip}
        -- ([xshift=1.7cm, yshift=0.08cm]pic cs:pushZero);
\end{tikzpicture}
\centering\footnotesize (d) License check function with multiple bypass points.
\end{minipage}
\caption{Dynamic analysis of StealC. The clean trace (c') shows the call to \texttt{check\_LicenseExpiration()}. A YARA rule (c'') can skip this call. The function (d) shows four potential bypass points: \colorbox{bypassClight}{Wret} at function entry, \colorbox{bypassAlight}{SetCF} to force the ``not expired'' branch, \colorbox{bypassBlight}{skip} on the ``expired'' jump, or \colorbox{skipcolorlight}{skip} on ExitProcess.}
\label{fig:dynamic_analysis}
\end{figure*}
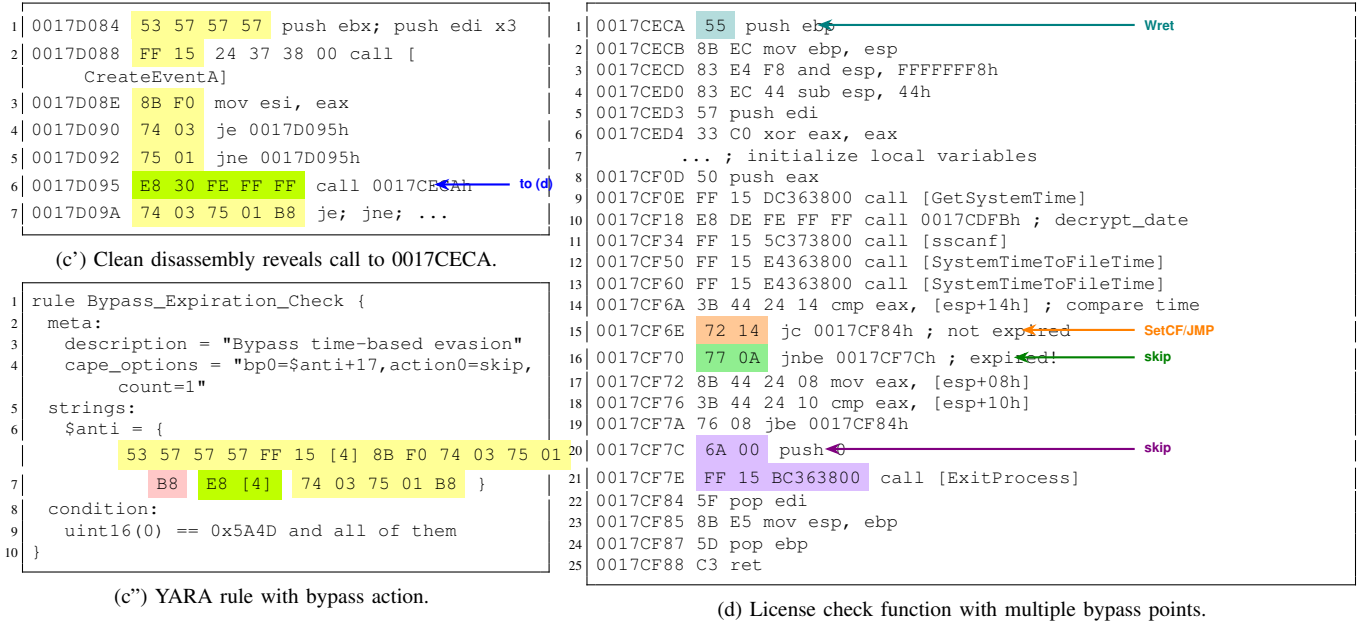
\subsection{Motivating Example: StealC Malware}
\label{sec:motivation}
We illustrate the evasion bypass challenge using StealC, an information stealer malware distributed via a decentralized MaaS model \cite{sekoia2023stealc}. Figure~\ref{fig:static_analysis}(a) shows the decompiled main function, which executes multiple anti-analysis checks before reaching its malicious payload. StealC checks RAM capacity at line 9 to detect VMs with limited memory. It then verifies the system language to avoid certain countries in line 10 and queries Windows Defender's emulator status in line 11. Notably, a license expiration is checked in line 25, where it terminates execution after a hardcoded date (see Figure~\ref{fig:static_analysis}(b)). 

When a compiled StealC binary is opened using a static analysis tool such as \textbf{IDA Pro}, analysis is complicated by the malware's anti-analysis techniques. Figure~\ref{fig:static_analysis}(c) shows how a junk byte (\texttt{B8}) corrupts IDA Pro's linear disassembly interpretation, causing instructions to be misinterpreted. Combined with opaque predicates\footnote{An opaque predicate is a condition that always evaluates to a constant outcome but is obfuscated to conceal its constant nature and hinder analysis.} (\texttt{74 03 75 01}—a \texttt{je} followed by a \texttt{jne} to the same target), the decompiler fails to recognize the call to \texttt{check\_LicenseExpiration()}. When this same binary is executed in a standard sandbox without bypass rules handled by the debugger, the malware terminates at the license check, hiding its C2 communication and data exfiltration capabilities. Figure~\ref{fig:dynamic_analysis}(c') shows the cleaned execution trace after removing the junk byte, revealing the actual call target. To bypass this evasion, an analyst must identify the check function, understand its logic, and craft a YARA rule with an appropriate action. Figure~\ref{fig:dynamic_analysis}(c'') shows an example of a rule that matches the anti-analysis patterns and skips the license check call.

Given the execution trace in Figure~\ref{fig:dynamic_analysis}(d), we observe multiple valid bypass strategies: (1) force return at function entry (\texttt{wret}), (2) manipulate flags to take the ``not expired'' branch, (3) skip the ``expired'' conditional jump, or (4) skip the \texttt{ExitProcess} call. When we provided this trace to an LLM without additional context, it generated multiple YARA rules targeting different bypass points. Some rules successfully bypassed the evasion in the sandbox, revealing the malware’s C2 communication behavior. However, LLM-generated rules also exhibited some problems: syntax errors, such as incorrect or malformed hex patterns; overly generic patterns that matched unintended locations and caused crashes; and incorrect offset calculations. This observation motivates our approach: \emph{LLMs have useful intuition about bypass strategies, but require validation and refinement through actual sandbox execution to produce working rules.}
\section{\tool: System Design}
\tool is designed as a fully automated framework that leverages LLM-based reasoning, rule sanitization, and iterative feedback from the sandbox to automatically detect evasion checkpoints, generate and validate YARA rules, and then execute the malware with the newly generated rules in a sandbox to bypass the evasion check.
\begin{figure}[b]
\centering \includegraphics[width=1\linewidth]{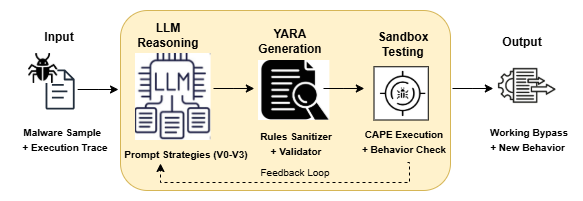}
\caption{Overview of \tool}
\label{fig:overview_simple}
\end{figure}

\textbf{As shown in \autoref{fig:overview_simple}, \tool workflow consists of three main components: (1) An LLM-guided YARA rule generator that receives and analyzes binary execution traces as input, identifies instruction sequences leading to potential early termination via \texttt{ExitProcess}, \texttt{TerminationProcess}}, or other related exit points, and uses LLM to generate YARA rules that specify the binary patterns used in the rules, the rule condition(s), and the corresponding bypass action(s) to take. (2) The second stage of \tool requires running a rule validation and repair sanitizer that applies format corrections and validation checks to ensure that generated YARA rules are syntactically correct and can execute properly in the sandbox, fixing common LLM formatting issues such as invalid wildcards. Rules that fail validation are rejected, and corrective feedback is provided to the LLM to regenerate revised rules. This validation and repair process continues until valid rules are produced. (3) The third component is an iterative sandbox execution and feedback engine that runs the sample while invoking the validated YARA rules within a sandbox. Success in this engine is measured by YARA rule hits and the progression of execution beyond the trace baseline. If the bypass fails, the engine analyzes the failure and uses this feedback to guide new LLM-based rules in subsequent iterations. 

\subsection{\textbf{LLM-Guided YARA Rule Generation}} \label{sec:trace} \tool uses LLMs to analyze execution traces and generate YARA rules with bypass actions through structured prompting. Recent studies have shown that LLMs trained on large corpora containing code and decompilation knowledge excel in code analysis \cite{ma2023lms,zhang2024detecting,khatun2025androbyte}, vulnerability analysis~\cite{cao2024llm, sheng2025llms, li2023llm}, patching~\cite{kim2025logs, kulsum2024case,tol2024zeroleak}, and code generation~\cite{shaikhelislamov2024llm, wang2023review, liu2024exploring}, suggesting that semantic and syntactic understanding of code and binaries can be achieved through prompting. While YARA rules are complex and require an understanding of binary traces to generate rules that correctly match evasion points, these models offer capabilities to validate binary hit points, reason about binary context, and produce correctly formatted rules. We integrate binary execution traces with marked exit points into the prompt to guide LLMs in generating valid YARA rules with actions for sandbox validation.

\begin{figure}[!t]
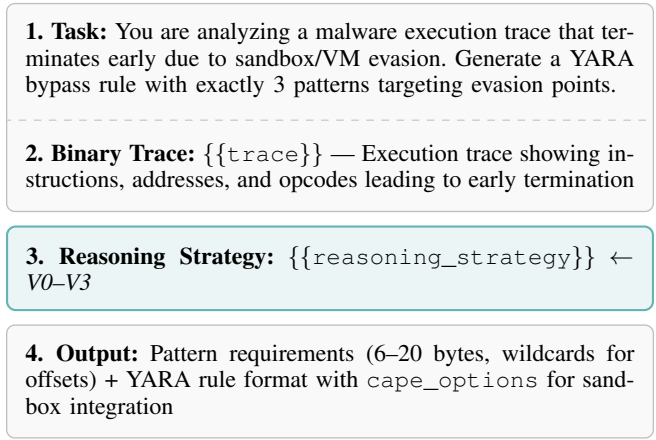

\centering
\small
\begin{tcolorbox}[left=4pt,right=4pt,top=4pt,bottom=4pt,colback=gray!5,colframe=gray!50,boxrule=0.5pt]
\textbf{1. Task:} You are analyzing a malware execution trace that terminates early due to sandbox/VM evasion. Generate a YARA bypass rule with exactly 3 patterns targeting evasion points.
\tcblower
\textbf{2. Binary Trace:} \texttt{\{\{trace\}\}} --- Execution trace showing instructions, addresses, and opcodes leading to early termination
\end{tcolorbox}
\begin{tcolorbox}[left=4pt,right=4pt,top=4pt,bottom=4pt,colback=teal!8,colframe=teal!60,boxrule=0.8pt]
\textbf{3. Reasoning Strategy:} \texttt{\{\{reasoning\_strategy\}\}} $\leftarrow$ \textit{V0--V3}
\end{tcolorbox}
\begin{tcolorbox}[left=4pt,right=4pt,top=4pt,bottom=4pt,colback=gray!5,colframe=gray!50,boxrule=0.5pt]
\textbf{4. Output:} Pattern requirements (6--20 bytes, wildcards for offsets) + YARA rule format with \texttt{cape\_options} for sandbox integration
\end{tcolorbox}
\caption{Prompt template structure. Components 1, 2, 4 are fixed; Component 3 varies across V0--V3.}
\label{fig:prompt_template}
\end{figure}

To generate effective YARA rules with proper bypass actions, \tool's prompt template combines four components as shown in Figure~\ref{fig:prompt_template}: The \textit{Binary Trace} provides the runtime execution trace of the target malware containing instruction sequences, memory addresses, and control-flow information. This data is inserted in the prompt via the \texttt{{{trace}}}. The \textit{Task} component specifies the analysis objective: to examine the provided execution trace, identify evasion-related instructions, and generate exactly \textit{n} distinct bypass patterns, producing a runnable debugger YARA rule with debugger action metadata that configures the sandbox to take the action specified in the generated rule upon execution at matched locations. This component also incorporates domain-specific knowledge about common evasion patterns, including \texttt{TEST+JE}, \texttt{CMP+JZ}, API result checks, timing checks, and instruction sequences that typically indicate program exit points. The \textit{Reasoning Strategy} defines the cognitive approach the LLM should employ, with four distinct strategies implemented as part of an ablation study: V0 (Zero-Shot)\cite{brown2020language}; V1 (Chain-of-Thought)~\cite{wei2022chain}; V2 (Counterfactual)~\cite{jin2023can}; and V3 (Adversarial)~\cite{perez2022red}.
The \textit{Pattern Requirements} enforce that each generated pattern must be 6--20 bytes long and specify that address bytes, jump offsets, and memory displacements must be replaced with wildcards (\texttt{??}), while opcode and register-encoding bytes are preserved to enable precise instruction matching. The \textit{Output Format} provides a complete YARA rule template with exact syntax, including the \texttt{strings} section with three pattern placeholders, the \texttt{condition} section, and the \texttt{meta} section with \texttt{cape\_options} as action (we used cape\_options for compatibility with evaluation sandbox, but can be modified if a different debugger is preferred).

\subsubsection{Reasoning Strategies}
\label{sec:reasoning-variants}
We evaluate four reasoning strategies with increasing analytical depth, where the key differentiator is how each strategy guides the LLM to analyze the execution trace and identify bypass points.
\paragraph{V0: Zero-Shot Baseline.}
This strategy provides direct task instruction without explicit reasoning guidance; the LLM receives only the task description and trace data:
\begin{tcolorbox}[left=2pt,right=2pt,top=2pt,bottom=2pt,colback=gray!5,colframe=gray!50,boxrule=0.5pt,title={\small V0: Zero-Shot}]
\small
Analyze the trace and generate a YARA bypass rule with exactly 3 different patterns targeting suspicious evasion points.
\end{tcolorbox}
\paragraph{V1: Chain-of-Thought.}
This strategy provides step-by-step reasoning that decomposes the analysis into sequential stages:
\begin{tcolorbox}[left=2pt,right=2pt,top=2pt,bottom=2pt,colback=gray!5,colframe=gray!50,boxrule=0.5pt,title={\small V1: Chain-of-Thought}]
\small
Think step-by-step: (1) The trace ends at an exit point---why did execution stop here? (2) What instruction or check CAUSED the program to decide to exit? (3) Where in the trace is that decision made? (4) What byte pattern can we match to skip that decision? Look for: function calls performing environment checks, conditional jumps leading to exit, comparisons determining program behavior.
\end{tcolorbox}
\paragraph{V2: Counterfactual Reasoning.}
In this reasoning strategy, a hypothesis is generated with systematic what-if analysis across multiple phases:
\begin{tcolorbox}[left=2pt,right=2pt,top=2pt,bottom=2pt,colback=gray!5,colframe=gray!50,boxrule=0.5pt,title={\small V2: Counterfactual}]
\small
\textbf{Phase 1 -- Hypothesis Generation:} For each suspicious instruction, ask: What is it checking? What happens if we skip it? Is this the root cause or a indicator? \textbf{Phase 2 -- Causal Analysis:} Trace backwards from ExitProcess, identify decision points where execution branches, find the critical check. \textbf{Phase 3 -- Counterfactual Scenarios:} For each candidate: (A) Skip instruction $\rightarrow$ state change? (B) Skip block $\rightarrow$ risks? (C) Modify return $\rightarrow$ confidence? \textbf{Phase 4 -- Evidence Scoring:} Rate 0--100 based on: matches known evasion (+30), early in trace (+20), leads to exit (+30), clean transition (+20).
\end{tcolorbox}
\paragraph{V3: Adversarial Reasoning.}
This last strategy provides a dual-perspective analysis from both attacker and defender viewpoints:
\begin{tcolorbox}[left=2pt,right=2pt,top=2pt,bottom=2pt,colback=gray!5,colframe=gray!50,boxrule=0.5pt,title={\small V3: Adversarial}]
\small
\textbf{Attacker's Perspective:} What is the malware trying to hide from (sandbox, VM, debugger)? What checks would an attacker implement? Where in the execution flow? \textbf{Defender's Perspective:} How can we detect these checks? What patterns reveal evasion attempts? Where is the minimum intervention point? \textbf{Decision Matrix:} For each candidate, score on: Necessity (is bypass required?), Sufficiency (is it enough?), Safety (side effects?), Robustness (works for variants?). Select highest-scoring candidate.
\end{tcolorbox}
\subsubsection{Pattern and Output Requirements}
All prompt versions share common pattern constraints embedded in the \textbf{output specification}. Patterns must be 6--20 bytes combining 2--3 consecutive instructions, wildcards (\texttt{??}) may only be used for addresses, offsets, and displacements, all patterns must come from the actual trace data, and each rule must include meta data section specifying breakpoint locations and debugger actions.
\begin{tcolorbox}[left=2pt,right=2pt,top=2pt,bottom=2pt,colback=gray!5,colframe=gray!50,boxrule=0.5pt,title={\small Output Format Sample}]
\small
\begin{verbatim}
rule Bypass_Sample {
  meta:
    cape_options = "bp0=$pattern0+0,action0=skip"
  strings:
    $pattern0 = { E8 ?? ?? ?? ?? 85 C0 74 ?? }
    $pattern1 = { FF 15 ?? ?? ?? ?? 85 C0 }
    $pattern2 = { 83 F8 01 0F 84 ?? ?? ?? ?? }
  condition: any of them
}
\end{verbatim}
\end{tcolorbox}
\subsection{Rule Validation and Repair Sanitization} \label{sec:validation}
While our prompt design aims to guide accurate rule generation, LLMs inherently exhibit hallucination and output inconsistencies. We observe that LLM-generated YARA rules frequently contain formatting errors and syntactic mistakes, likely due to the models' exposure to diverse programming language patterns during training. To ensure that generated rules are deployable within the sandbox environment, \tool implements a three-stage validation pipeline consisting of auto-sanitization, syntactic validation, and semantic validation. This pipeline enforces rule correctness and executability before rules are passed to the analysis engine, preventing malformed rules from disrupting iterative sandbox execution.

\subsubsection{Error Pattern Analysis}
To develop effective sanitization strategies, we conducted an empirical analysis of rule generation errors across multiple LLMs by deploying smaller, efficient models \textbf{LLama3.1-8B, Qwen3-8B, DeepSeek-R1-7B, Gemma-12B)} to generate bypass rules for 200 malware samples and collecting all failed generation attempts. We then employed a larger, more capable model (Claude Opus 4.1~\cite{anthropic2025opus41}) to analyze these failure cases and identify recurring error patterns. This analysis revealed that different models exhibit distinct error signatures correlated with their training data composition: models with substantial SQL or Lua training data generate \texttt{-{}-} style comments; models heavily trained on Python or shell scripts produce \texttt{\#} style comments; models trained on mixed programming corpora confuse hex byte notation with assembly mnemonics, generating patterns like \texttt{\{PUSH EAX CALL EDX\}} instead of valid hex bytes; and models with limited exposure to YARA syntax produce malformed wildcards (\texttt{???}, \texttt{????}) or incorrect pattern delimiters. Based on this analysis, the advanced model generated targeted fix strategies for each error category, leveraging the superior reasoning capabilities of larger models to compensate for the syntactic limitations of smaller, more cost-effective models used in the framework. Table~\ref{tab:sanitization} summarizes some of the identified error patterns and corresponding auto-correction rules.
\begin{table}[t]
\centering
\footnotesize
\caption{Auto-sanitization rules for LLM-generated patterns}
\label{tab:sanitization}
\begin{tabular}{@{}lll@{}}
\toprule
\textbf{Error Pattern} & \textbf{Invalid} & \textbf{Corrected} \\
\midrule
SQL/Lua comments & \texttt{-{}-comment} & \texttt{//comment} \\
Python comments & \texttt{\#comment} & \texttt{//comment} \\
Malformed wildcards & \texttt{???} & \texttt{??} \\
Hex prefix & \texttt{0xFF 0x15} & \texttt{FF 15} \\
Quoted strings & \texttt{"E8 ??"} & \texttt{\{E8 ??\}} \\
Missing spaces & \texttt{E8??83F8} & \texttt{E8 ?? 83 F8} \\
Inline annotations & \texttt{\{74 (je)\}} & \texttt{\{74\}} \\
\bottomrule
\end{tabular}
\end{table}

\subsubsection{Auto-Sanitization}
Based on the error patterns identified above, \tool applies deterministic transformations to correct formatting issues without requiring LLM re-generation, which often results in the model repeatedly producing identical errors As shown in the Figure \ref{fig:validation_pipeline}, the sanitization module processes each rule through a series of regex-based transformations that convert invalid syntax to YARA-compliant format. To complement formatting corrections and ensure rule executability, \tool automatically injects the action metadata field when absent, specifying breakpoint locations and debugger actions required to bypass evasion checks at runtime.

\usetikzlibrary{positioning, arrows.meta, shapes.geometric}
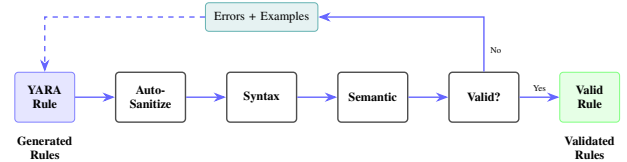
\begin{figure}[h]
\centering
\resizebox{0.95\columnwidth}{!}{%
\begin{tikzpicture}[
    node distance=0.8cm,
    arrow/.style={-{Stealth[length=2.5mm]}, thick, draw=blue!60},
    dasharrow/.style={-{Stealth[length=2.5mm]}, thick, draw=blue!60, dashed},
    stage/.style={draw=black!70, thick, minimum width=1.4cm, minimum height=1cm, fill=white, font=\scriptsize\bfseries, align=center, rounded corners=2pt},
    doc/.style={draw=blue!50, fill=blue!10, minimum width=1.2cm, minimum height=1cm, font=\scriptsize\bfseries, align=center, rounded corners=2pt},
    feedback/.style={draw=teal!60, fill=teal!10, rounded corners=2pt, font=\scriptsize, align=center, inner sep=5pt},
    label/.style={font=\scriptsize\bfseries, align=center},
]
\node[doc] (in) {YARA\\Rule};
\node[label, below=0.15cm of in] {Generated\\Rules};
\node[stage, right=of in] (s1) {Auto-\\Sanitize};
\node[stage, right=of s1] (s2) {Syntax};
\node[stage, right=of s2] (s3) {Semantic};
\node[stage, right=of s3] (s4) {Valid?};
\node[doc, fill=green!10, draw=green!50, right=of s4] (out) {Valid\\Rule};
\node[label, below=0.15cm of out] {Validated\\Rules};
\node[feedback, above=0.8cm of s2] (retry) {Errors + Examples};
\draw[arrow] (in) -- (s1);
\draw[arrow] (s1) -- (s2);
\draw[arrow] (s2) -- (s3);
\draw[arrow] (s3) -- (s4);
\draw[arrow] (s4) -- node[above, font=\tiny] {Yes} (out);
\draw[arrow] (s4.north) |- node[pos=0.2, right, font=\tiny] {No} (retry.east);
\draw[dasharrow] (retry.west) -| (in.north);
\end{tikzpicture}
}%
\caption{YARA rule auto-sanitization with self-correction feedback loop.}
\label{fig:validation_pipeline}
\end{figure}

\subsubsection{Syntactic Validation}
After sanitization, \tool validates structural correctness against YARA grammar. To be executable by the sandbox debugger, a valid bypass rule must contain: (1) a rule declaration with an identifier, (2) a \texttt{strings} section defining hex byte patterns, (3) a \texttt{condition} section specifying match logic, and (4) metadata for sandbox integration (\texttt{cape\_options} for our implementation). \tool specifically rejects patterns containing assembly mnemonics (e.g., \texttt{PUSH}, \texttt{CALL}) or register names (e.g., \texttt{EAX}, \texttt{EDX}), as these indicate the LLM confused hex byte representation with assembly text; for such cases, the validation error message provides explicit feedback on converting assembly to hex opcodes, enabling effective self-correction in retry attempts.

\subsubsection{Semantic Validation}
Syntactically correct rules may still fail to achieve bypass objectives if patterns are insufficiently specific, as the debugger sets breakpoints by matching binary patterns, and overly generic patterns may match unintended code locations across the binary. Through sandbox deployment experiments, we identified semantic constraints critical for effective bypass rules. Each pattern must contain 6--20 bytes, as shorter patterns risk matching unintended code locations while longer patterns reduce portability across malware variants. Patterns exceeding 80\% wildcard bytes provide insufficient discrimination and match nearly arbitrary byte sequences. Standalone API call patterns (\texttt{FF 15 ?? ?? ?? ??} for indirect calls, \texttt{E8 ?? ?? ?? ??} for direct calls) require additional surrounding context bytes to avoid overly broad matches. All patterns within a rule must be distinct to maximize coverage of different evasion decision points and avoid conflicting actions on the same breakpoint, which can cause execution crashes.

\subsubsection{Self-Correction with Error Feedback}
When validation fails, \tool constructs an augmented prompt containing the specific errors, the failed rule, and correction examples derived from our error pattern analysis. The retry prompt includes enumerated validation errors with explanations, the original failed rule for reference, concrete before/after fix examples for each error type, and the original trace data to maintain context. \tool attempts up to three correction iterations before aborting the generation attempt. To avoid over-constraining the iterative refinement process, semantic checks for context-sensitive patterns are enforced only on the initial attempt; subsequent iterations permit these patterns, allowing the model flexibility to make informed trade-offs. The complete validation and sanitization algorithm is presented in Appendix~\ref{appendix:validation-algorithm}, Algorithm~\ref{alg:validation}.
\subsection{Sandbox Execution and Feedback Engine} \label{sec:sandbox}
\tool validates bypass rules through actual malware execution in a sandbox, measuring success based on observable behavioral changes rather than static metrics. This empirical validation ensures that generated rules bypass evasion checkpoints in practice.
\subsubsection{Baseline Establishment}
Before evaluating any bypass rule, \tool establishes a behavioral baseline by executing the malware sample without any YARA rules deployed. This baseline captures the sample's default behavior when its evasion checks succeed---typically resulting in early termination with minimal observable activity. 
The baseline execution records three key metrics: the set of triggered behavioral signatures, the API calls log, and the overall maliciousness score assigned by the sandbox, serving as the reference point against which all subsequent bypass attempts are compared. To ensure robust baseline collection, \tool executes the sample at least four times with VM recovery between attempts, selecting the execution with the highest signature count as the authoritative baseline to account for non-deterministic sandbox behavior.
\subsubsection{Rule Deployment and Execution}
Once the baseline is established, \tool deploys candidate YARA rules to the CAPE sandbox for validation. As shown in Appendix~\ref{appendix:Sandbox-algorithm}, Algorithm~\ref{alg:sandbox}, the deployment process uploads the generated YARA rule to the sandbox's rule directory, where the YaraScan module of the sandbox processes it during analysis. The rule's metadata instructs the sandbox to set breakpoints at matched pattern instruction location and take debugger actions as specified in the rule. After rule deployment, \tool submits the malware sample for analysis, during which the sandbox monitors whether the YARA rule matches any location (\textit{rule hit}) and records all behavioral signatures triggered during the extended execution. The sandbox also captures debugger logs when crashes occur, providing detailed information about the instruction that was matched and new execution flow. These logs additionally provide the faulting instruction trace, for example the debugger\_log detail sample show during stage 3 in Appendix~\ref{appendix:workflowexample}, figure~\ref{fig:case_study}.

\subsubsection{Success Criteria}
\tool defines bypass success through a conjunction of two conditions: the YARA pattern must match the malware instructions during execution (\textit{rule hit}), confirming that the bypass mechanism was activated, and the execution must reveal at least one new behavioral signature not present in the baseline, indicating that previously hidden functionality was exposed. This dual requirement prevents false positives from two failure modes: a rule that matches but produces no new behaviors likely targeted the wrong location, while new signatures without a rule hit suggest coincidental result rather than successful bypass. \tool classifies validation outcomes into four categories: \textit{Success} (rule matched and new signatures discovered), \textit{Partial} (rule matched but no new signatures), \textit{Crashed} (rule caused program termination), and \textit{Failed} (rule did not match). Algorithm~\ref{alg:sandbox} presents the complete validation procedure.

\subsubsection{Feedback-Driven Rule Iteration} \label{sec:evolution}
When initial bypass attempts fail, \tool employs an iterative refinement process that leverages structured feedback to guide LLM-based rule updates. Rather than regenerating rules from scratch, \tool analyzes each failure to construct targeted feedback that informs subsequent attempts. For \textit{Crashed} failures, \tool extracts the debugger log showing the matched instruction and its context, then guides the LLM to select a safer nearby instruction as the bypass target. For \textit{Partial} failures where the pattern matched a potentially wrong location, \tool instructs the LLM to target instructions closer to the evasion checkpoint. For \textit{Failed} outcomes where the pattern did not match, \tool suggests extending the pattern with additional context bytes or selecting alternative instruction sequences from the trace. The debugger log proves particularly valuable for crash failures, as it provides the exact instruction that was matched or where the action was taken, allowing the LLM to determine whether the crash resulted from a pattern that was too generic or from targeting an instruction that was unsafe to be manipulated by the debugger action, or we need to choose a different break-point. For iterative rule generation following crashes, partial successes, or failed executions, \tool constructs revised prompts that incorporate the original execution trace for reference, the history of previous attempts and their outcomes, the current best-performing rule, and targeted feedback derived from the most recent non-successful execution. This combination enables the LLM to learn from prior failures supporting systematic convergence rather than repeated trial-and-error. Algorithm~\ref{alg:evolution} in Appendix ~\ref{appendix:feedback-algorithm} describes this iterative loop.
A detailed StealC example illustrating how ABLE workflow handles evasion is provided in Appendix \ref{appendix:workflowexample}.

\section{Implementation}
We implemented \tool as a multi-component framework in Python, consisting of $\approx 11,000$ lines of code. Each component is built as a plugin and can be inserted or removed as needed, which significantly contributes to the system’s effectiveness and provides flexibility for integrating future components. Additionally, \tool operates as a lightweight post-processing integration, in contrast to runtime integration solutions that require substantially more complex engineering effort, making it portable across different sandbox analysis platforms.

To support the generation of YARA bypass rules from execution traces, we use the CAPE sandbox~\cite{cape2024} to collect execution traces and validate generated rules. CAPE is one of the most widelt used malware analysis sandbox that supports YARA-based breakpoint injection, enabling us to test bypass rules using available control-flow manipulation commands, such as skip and jump. Specifically, we first collect execution traces leading to early termination. We then leverage CAPE’s \texttt{cape\_options} metadata to deploy bypass rules that set breakpoints at matched patterns and manipulate instruction control flow as specified by the generated rule. Behavioral signatures triggered during bypassed execution, when compared against baseline runs without rules, indicate whether the YARA action successfully bypasses the evasion check. For our proof of concept, we set the maximum number of iterations to three. After three attempts, the tool terminates the feedback-driven iterative loop. Our empirical analysis in Section~\ref{sec:eval} explains the rationale for selecting this iteration budget.

The sandbox engine is implemented using an SSH-based controller that manages YARA rule deployment, sample submission, and result retrieval from CAPE VMs. The controller handles connection failures through retry logic and can trigger VM recovery (snapshot restore). For multi-VM deployments, the controller supports parallel submission to improve throughput, provide redundancy against individual VM failures, and accelerate experimentation.
\section{Evaluation}\label{sec:eval}
We evaluate \tool through a combination of quantitative and quantitative analysis to assess its effectiveness, design contributions, and practical impact on malware analysis. Specifically, we address the following research questions:

\noindent\textbf{RQ1}: How effective is \tool for real-world malware analysis? We compare sandbox signatures obtained using \tool’s generated bypass rules against baseline executions performed without bypass rules.

\noindent\textbf{RQ2}: How do individual design components of \tool contribute to bypass success?
We conduct an ablation study to quantify the contribution of the rule sanitizer, LLM model choice, reasoning strategy, and iterative feedback on overall evasion bypass effectiveness.

\noindent\textbf{RQ3}: How effective is \tool at discovering new behavioral signatures and malware family classifications? We evaluate the types of signatures exposed and the number of malware families identified relative to baseline sandbox analysis.

\subsection{Data Collection:} 
We leveraged MalwareBazaar~\cite{malwarebazaar_sample_caf}, a platform developed by abuse.ch~\cite{abusech} in collaboration with Spamhaus~\cite{spamhaus}, accessing its daily malware sample collection spanning January 2020 to December 2024. From this corpus, we selected 334 malware samples marked as exhibiting some form of sandbox evasion by VirusTotal~\cite{virustotal_sample_caf}, JoeSandbox \cite{joesandbox2024} and CAPA~\cite{capa2024}, including indicators of anti-debugging, anti-virtualization, and early termination logic. These annotations serve as widely used, independent signals of evasive behavior rather than exhaustive ground truth.
The dataset primarily consists of Windows PE malware, dominated by native Windows PE32 GUI binaries (77.8\%), followed by .NET assemblies (18.3\%), Console-based and UPX packed samples represent a small minority (3.0\% and 0.9\%, respectively), aligning with the scope of contemporary sandbox-based analysis. This selection enables evaluation of \tool under realistic conditions where evasive behavior is expected but not fully observable under baseline execution.

\subsection{Model, Reasoning Strategy and Iteration}
We evaluate \tool using four open-weight large language models for bypass rule generation: Qwen3-8B\cite{qwen2025qwen3}, LLama~3.1-8B\cite{dubey2024llama}, Gemma~3-12B\cite{gemmateam2024gemma}, and DeepSeek-R1-7B\cite{deepseekai2025deepseekr1}. These models were selected to represent a diverse set of architectures, training objectives, and reasoning capabilities, while remaining deployable in offline or controlled analysis environments commonly used for malware research.
All models operate in a local inference setting and are prompted using identical templates to isolate the impact of model reasoning behavior from prompt design. We select models in the 7B–12B parameter range to balance inference cost with reasoning capability, enabling repeated iterative execution within the feedback loop. This choice demonstrate realistic deployment constraints for automated sandbox environment, where scalability and analysis capability are critical, smaller models facilitate parallel deployment across multiple concurrent sandbox instances in isolation environment.
The selected models differ in their architecture relevant to this task. Qwen3-8B and LLama~3.1-8B represent strong general-purpose instruct models. Gemma~3-12B provides increased capacity for structured reasoning, while DeepSeek-R1-7B emphasizes explicit reasoning and multi-step analysis. Evaluating these models allows us to assess how different reasoning styles contribute to bypass success and rule quality, and to determine whether combining models through ensembling provides improved effectiveness under identical execution conditions.

Each model was evaluated using the four prompt strategies: V0 (Zero-shot), V1 (Chain-of-Thought), V2 (Counterfactual), and V3 (Adversarial), as described in Section~\ref{sec:trace}. This results in 16 distinct model–prompt configurations. For each configuration, \tool performs up to three sandbox iterations (\texttt{iter0}, \texttt{iter1}, \texttt{iter2}), with feedback-driven rule refinement between iterations. 

\subsection{System Setup and Analysis Configuration}\label{5.3}
Our system setup employs nested virtualization for CAPEv2 to ensure the isolation of the analysis host, trace and analysis log collection. Our hardware configuration consists of a host machine equipped with a 24-core (13th Gen Intel Core i9-13900KF) CPU, 64 GB memory, and an NVIDIA GeForce RTX 4090 GPU (24 GB VRAM) for serving ollama with temperature set to 0.7 and a maximum context length of 128K tokens. We use VMware Workstation Pro 17.0 to create an isolated CAPEv2 running environment. The outer VM is allocated 8 virtual CPU cores and 8 GB of memory, running Ubuntu 22.04 LTS. And the victim VM, nested within the Ubuntu VM, runs Win10 x86 with 8 virtual CPU cores and 8 GB of memory.

\textbf{We note that, while \tool is evaluated on 334 samples, our analysis ran a total of 13,778 sandbox executions. Each sandbox execution is subject to a maximum analysis time of 200 seconds.}

\begin{table*}[t]
\caption{Overall results for \tool-generated bypass rules on 334 malware samples (excluding crashed)}\label{tab:overall-results-no-crash}
\centering
\begin{threeparttable}
\setlength{\tabcolsep}{5pt}
\footnotesize
\resizebox{\textwidth}{!}{%
\begin{tabular}{l|cccc|cccc|cccc|ccc|cc}
\toprule[1.5pt]
& \multicolumn{4}{c|}{\textbf{@iter0}} & \multicolumn{4}{c|}{\textbf{@iter1}} & \multicolumn{4}{c|}{\textbf{@iter2}} & \multicolumn{3}{c|}{\textbf{Total}} & \multicolumn{2}{c}{\textbf{Success}} \\
\midrule[0.4pt]
\textbf{Model} & \textbf{V0} & \textbf{V1} & \textbf{V2} & \textbf{V3} & \textbf{V0} & \textbf{V1} & \textbf{V2} & \textbf{V3} & \textbf{V0} & \textbf{V1} & \textbf{V2} & \textbf{V3} & \textbf{iter0} & \textbf{iter1} & \textbf{iter2} & \textbf{Unique} & \textbf{Shared} \\
\midrule[0.8pt]
DeepSeek-R1 (7B) & \textcolor{red}{39} & 26 & 26 & 30 & 56 & 40 & 47 & 48 & 67 & 55 & 60 & 67 & 81 & 104 & 123 & 11 & 112 \\
Gemma3 (12B) & \textcolor{red}{39} & \textcolor{red}{45} & \textcolor{red}{38} & \textcolor{red}{50} & \textcolor{red}{69} & \textcolor{red}{75} & \textcolor{red}{69} & \textcolor{red}{76} & 81 & 88 & \textcolor{red}{91} & \textcolor{red}{91} & \textcolor{red}{97} & 134 & 144 & 32 & 112 \\
LLaMA3.1 (8B) & 27 & 30 & 18 & 24 & 43 & 43 & 39 & 38 & 49 & 54 & 48 & 47 & 59 & 82 & 93 & 10 & 83 \\
Qwen3 (8B) & 28 & 33 & 26 & 34 & 67 & 67 & 62 & 60 & \textcolor{red}{89} & \textcolor{red}{92} & 87 & 79 & 80 & \textcolor{red}{147} & \textcolor{red}{174} & \textcolor{red}{49} & \textcolor{red}{125} \\
\midrule[0.8pt]
\textbf{All} & \textbf{102} & \textbf{107} & \textbf{88} & \textbf{106} & \textbf{162} & \textbf{157} & \textbf{157} & \textbf{159} & \textbf{187} & \textbf{184} & \textbf{188} & \textbf{187} & \textbf{186} & \textbf{251} & \textbf{264} & \textbf{102}$^\dagger$ & \textbf{25}$^\ddagger$ \\
\bottomrule[1.5pt]
\end{tabular}}
\begin{tablenotes}
\item
\textbf{V0}=Zero-shot; \textbf{V1}=CoT; \textbf{V2}=Counterfactual; \textbf{V3}=Adversarial.
@iter1/@iter2 show cumulative success (samples solved by that iteration). \textbf{Excludes crashed samples.}
\item
\textbf{Unique} = samples solved \textit{only} by just one model; \textbf{Shared} = samples also solved by other models.
$^\dagger$102 samples solved by exactly one model. $^\ddagger$25 samples solved by all four models.
\end{tablenotes}
\end{threeparttable}
\end{table*}

\subsection{RQ1: Effectiveness of \tool's Framework}
To evaluate the effectiveness of our framework, we compare sandbox signatures obtained using generated bypass rules against baseline executions performed without rules. We define bypass success conservatively: crule is considered successful only if (1) it compiles correctly and produces rule hits recorded in sandbox logs, (2) it exposes new behavioral signatures not observed during baseline execution, and (3) the sandbox execution completes without crashing.

Table~\ref{tab:overall-results-no-crash} summarizes the results. At iteration~0 (without feedback), individual model–prompt configurations successfully bypass between 18 and 50 samples (5.4\%–15.0\%). The best single configuration in this first try (iteration~0) is Gemma3-12B with the Adversarial reasoning strategy, achieving bypass for 50 out of 334 samples ($\approx 15\%$). While combining all 16 configurations covers 186 unique samples (55.7\%), indicating that \tool can automatically generate bypass rules for over half of the samples in the first iteration using four prompt strategies across four lightweight models. In contrast, the best overall model–reasoning combination across all three iterations is Qwen3-8B with the Chain-of-Thought strategy, which achieves bypass for 92 samples by the second iteration ($\approx 27.5\%$). Ensembling all the models, \tool achieves 264 out of 334 samples (79.0\%) bypass success across all configurations. Showing that iterative refinement contributes an additional 78 samples, increasing coverage from 186 --> 251 --> 264 samples (79.0\%) by iteration~2. We note that this result excludes 11 samples in which the sandbox crashed but still reported new signatures. Additionally, our results across the 16 configurations also shows that only 25 samples succeed across all four models (Shared Column), while 102 samples succeed with exactly one model (Unique Column). Overall, these results demonstrate that both model–prompt diversity and iterative refinement substantially improve bypass coverage. While individual configurations benefit from feedback, no single model–reasoning strategy approaches the coverage achieved by ensembling multiple configurations. The strong performance gap between the best individual configuration (27.5\%) and the ensemble (79.0\%) indicates that different models capture complementary evasion strategies across samples. 


\begin{tcolorbox}[left=2pt,right=2pt,top=2pt,bottom=2pt,colback=gray!5,colframe=gray!50,boxrule=0.5pt,title={\small RQ1: Takeaway}]
\small
\tool substantially improves dynamic malware analysis by reliably bypassing sandbox evasion mechanisms at scale. Compared to baseline sandbox execution without bypass rules, \tool exposes previously hidden behaviors in 79.0\% of evasive samples, significantly expanding observable signatures and malware family coverage. Overall, while LLMs enable effective YARA rule generation, ensembling models significantly outperforms individual configurations under identical execution conditions.
\end{tcolorbox}

\subsection{RQ2: \tool Design Component Analysis}
To understand why \tool is effective, we conduct a systematic ablation study analyzing the contribution of its key design components. Specifically, we evaluate the impact of the rule sanitizer, LLM model choice, reasoning strategy, and iterative feedback loop on evasion bypass success. By isolating and disabling individual components, we quantify how each design decision influences rule validity, convergence behavior, and overall bypass coverage.

\begin{table}[h]
\caption{Sanitizer validation retry rates by component}\label{tab:sanitizer}
\centering
\footnotesize
\setlength{\tabcolsep}{3pt}
\begin{threeparttable}
\begin{tabular}{l|l|r|r|r}
\toprule
\textbf{Component} & \textbf{Config} & \textbf{Calls} & \textbf{Retries} & \textbf{Rate} \\
\midrule
Model & DeepSeek-R1(7B) & 3,631 & 2,221 & 61.2\% \\
Model & Gemma3(12B) & 3,908 & 1,743 & 44.6\% \\
Model & LLama3.1(8B) & 2,966 & 717 & \textcolor{red}{24.2}\% \\
Model & Qwen3(8B) & 3,273 & 956 & 29.2\% \\
\midrule
Prompt & V0 (Zero-shot) & 3,489 & 1,290 & \textcolor{red}{37.0}\% \\
Prompt & V1 (CoT) & 3,425 & 1,494 & 43.6\% \\
Prompt & V2 (Counterfactual) & 3,361 & 1,421 & 42.3\% \\
Prompt & V3 (Adversarial) & 3,503 & 1,432 & 40.9\% \\
\midrule
Iteration & Iteration 0 & 8,820 & 5,041 & 57.2\% \\
Iteration & Iteration 1 & 4,720 & 596 & 12.6\% \\
Iteration & Iteration 2 & 238 & 0 & \textcolor{red}{0.0}\% \\
\bottomrule
\end{tabular}
\begin{tablenotes}
\footnotesize
\item Rate = percentage of LLM outputs that failed initial validation and required rule repair/correction.
\item Model/Prompt rates aggregated across all iterations; Iteration rates aggregated across all models and prompts.
\end{tablenotes}
\end{threeparttable}
\end{table}
\subsubsection{Sanitizer Contribution}
The sanitizer validates LLM-generated YARA rules and provides error feedback and automated correction. Quantitatively, for the 13,778 executions, the sanitizer detected a total of 9,284 validation errors across all experiments and produced 5,262 validated LLM-generated rules. An ablation that removes sanitization would therefore discard or invalidate the majority of candidate rules before runtime evaluation, effectively reducing the search space and decreasing the probability of successful evasion bypass. When mapped to individual malware samples, removing the sanitizer’s retry mechanism reduces bypass success from 264 to 232 samples—a loss of 32 samples (12.1\%).

An additional quantitative metric, shown in Table~\ref{tab:sanitizer}, shows substantial variation in retry rates across the rest of the variables - model choice, prompt strategy, and iteration. Across models, DeepSeek-R1 has the highest retry rate (61.2\%), while LLaMA~3.1 achieves the lowest (24.2\%). However, this variation does not directly correlate with final bypass success (Sandbox Feedback Loop). In contrast, DeepSeek-R1 requires more frequent correction because its \textbf{<thinking>} tag content often lacks sufficient binary-specific reasoning, which can interfere with the generated rule content. Nevertheless, the reasoning traces remain useful for iterative refinement, and DeepSeek-R1 ultimately succeeds on a larger number of samples despite its higher retry rate. The prompt strategy exhibits the smallest variation in retry rate, ranging from 37.0\% (V0) to 43.6\% (V1). The widest variation in retry rate occurs across iterations. This rate decreases from 57.2\% at iteration~0 to 12.6\% at iteration~1, and then to 0\% at iteration~2. This reduction occurs because rule updates task to the LLM in subsequent iterations integrate the sandbox execution feedback, including detailed debugger logs that capture the execution trace around the inserted breakpoints and the rules from prior failed attempts. This feedback enables the LLM to generate rules that more closely align with actual malware sandbox evasion behavior, rather than relying solely on the initial execution trace. We note that this empirical evaluation provides the rationale for selecting three as the iteration budget, as no significant improvement in rule generation is observed beyond three iterations 

Finally, while the sanitizer is designed to filter and fix syntax and formatting errors across rules generated by different models, as described in Section~\ref{sec:validation}, the error distribution reveals common LLM limitations in YARA rule generation. Specifically, \num{5815} errors (62.6\%) involve patterns shorter than the minimum length threshold, 680 errors (7.3\%) involve duplicate patterns, and the remaining errors are from structural issues, including invalid rule formats, missing CAPE options, and other simple YARA syntax errors.
\begin{table}[h]
\caption{Ablation study: Impact of removing each component}\label{tab:ablation}
\centering
\footnotesize
\setlength{\tabcolsep}{3pt}
\begin{threeparttable}
\begin{tabular}{l|l|r|r|r}
\toprule
\textbf{Component} & \textbf{Removed} & \textbf{Keep} & \textbf{Loss} & \textbf{\%} \\
\midrule
\textit{Full System}& \textit{--} & \textit{264} & \textit{--} & \textit{--} \\
\midrule
Sanitizer$^\star$ & Retry & 232 & $-$32 & $-$12.1\% \\
\midrule
Model & DeepSeek-R1 (7B) & 253 & $-$11 & $-$4.2\% \\
Model & Gemma3 (12B) & 232 & $-$32 & $-$12.1\% \\
Model & LLama3.1 (8B) & 254 & $-$10 & $-$3.8\% \\
Model & Qwen3 (8B) & 215 & $-$49 & \textcolor{red}{$-$18.6}\% \\
\midrule
Prompt & V0 (Zero-shot) & 246 & $-$18 & \textcolor{red}{$-$6.8}\% \\
Prompt & V1 (CoT) & 253 & $-$11 & $-$4.2\% \\
Prompt & V2 (Counterfactual) & 253 & $-$11 & $-$4.2\% \\
Prompt & V3 (Adversarial) & 255 & $-$9 & $-$3.4\% \\
\midrule
Iteration & iter2 & 251 & $-$13 & $-$4.9\% \\
Iteration & iter1 + iter2 & 186 & $-$78 & \textcolor{red}{$-$29.5}\% \\
\bottomrule
\end{tabular}
\end{threeparttable}
\end{table}
\subsubsection{Model, Prompt, and Iteration Contributions}
Table~\ref{tab:ablation} presents the ablation study results. Among the evaluated models, Qwen3 provides the largest individual contribution: removing it reduces success from 264 to 215 samples, corresponding to an 18.6\% loss. The remaining models contribute smaller but consistent gains, with losses ranging from 3.8\% to 12.1\% when removed. Prompt variations contribute more modestly, with success reductions ranging from 3.4\% (V3) to 6.8\% (V0) when individual strategies are removed. The relatively uniform impact across prompt strategies suggests that prompt diversity provides incremental benefit by covering different reasoning approaches, rather than any single prompt dominating performance. Iterative refinement provides the largest cumulative contribution. Removing iteration~2 alone results in a 4.9\% loss, while removing both iterations~1 and~2 leads to a 29.5\% reduction, decreasing the number of successful samples from 264 to 186. This result demonstrates that iterative refinement based on sandbox execution feedback, together with repeated rule sanitization, is essential for designing valid LLM-generated bypass rules and achieving improved analysis progression.

\begin{tcolorbox}[left=2pt,right=2pt,top=2pt,bottom=2pt,colback=gray!5,colframe=gray!50,boxrule=0.5pt,title={\small RQ2: Takeaway}]
\small
The ablation study reveals that \tool’s effectiveness is driven primarily by execution-grounded iteration rather than any single LLM, prompt strategy, or heuristic. Iterative refinement based on sandbox feedback provides the largest performance gains. Model diversity contributes the next most significant benefit, while prompt variation offers smaller but consistent improvements by covering diverse reasoning approaches. 
\end{tcolorbox}

\subsection{RQ3: \tool Qualitative Analysis}
\subsubsection{Signature Type Analysis}
Across the 264 samples that resulted in success, ABLE identified 82 distinct hidden signatures (per CAPEv2 signature specifications). Overall, \texttt{queries\_keyboard\_layout} is the most frequently observed signature, appearing in 109 samples, followed by \texttt{terminate\_remote\_process} in 105 samples, and \texttt{antisanbox\_unhook} in 65 samples. Compared with baseline analysis, these signatures reveal behaviors previously hidden. Additionally, out of the 82 hidden signatures, four were completely new {not found in the CAPEv2 baseline result signature set)—\texttt{persistence\_ads}, \texttt{uses\_remote\_desktop\_session}, \texttt{network\_anomaly}, and \texttt{anomalous\_deletefile} —and were observed across 19 malware samples. A complete list of all detected malware signatures and their distribution in our sample is provided in Appendices \ref{appendix:newsig},{}\ref{appendix:newsig2} \& \ref{appendix:newsig3}.

\subsubsection{Analysis Case Studies} 
\textbf{Case 1 - e536afc7f63611d1bbea4305f958661e}: This sample is identified as a RedLine or Amadey malware installer. After bypassing a time-check evasion mechanism, the malware reveals a previously hidden self-deletion behavior (\texttt{anomalous\_deletefile signature}). Specifically, the sample deletes more than ten suspicious executable files, including \texttt{mnolyk.exe.job}, which does not appear in publicly accessible reports from MalwareBazaar~\cite{malwarebazaar_sample_caf} or VirusTotal~\cite{virustotal_sample_caf}.
In the AnyRun analysis report~\cite{anyrun_e536afc7}, the sandbox detects six file-deletion events; however, it does not observe access to \texttt{mnolyk.exe.job}. Additionally, the malware causes an abnormal shutdown of the sandbox victim window, potentially concealing further observable behavior. The presence of the \texttt{mnolyk.exe.job} artifact indicates that the malware deploys a scheduled task associated with the Amadey malware loader, a behavior relevant to persistence and execution control.\\

\textbf{Case 2 - 584e1b772abf47e40b94652be9117be8}: 
This sample is identified as SystemBC malware. After bypassing its evasion mechanism—which hides the network activity (\texttt{network\_anomaly} signature not documented in existing public analyses)—the malware reveals its subsequent operational behavior, including active scanning and connection attempts to malicious IP addresses originating from Japan.
Following evasion bypass, we successfully extract the malware’s configuration, revealing the following network indicators:SystemBC {"HOST1": "210.16.67.250", "HOST2": "192.168.1.28", "PORT1": "3000"}. These indicators are consistent with reports from Zenbox as aggregated in VirusTotal~\cite{virustotal_sample_caf} and corroborated by analysis from Joe Sandbox~\cite{joesandbox2024}. However, under CAPEv2 baseline execution without evasion bypass, this network behavior remains hidden and is not observable during sandbox analysis.\\

\textbf{Case 3 - 40c37a1b79cfcf7f26d43c7f5209cf55}: FormBook employs a unique RunPE-based crypter which was initially referred to as the “Babushka Crypter” by Insidemalware. After running this sample through \tool, we observe previously hidden behavior in which the malware invokes \texttt{C:\textbackslash Windows\textbackslash SysWOW64\textbackslash mstsc.exe} via the command line, attempting to initiate a Remote Desktop session (\texttt{uses\_remote\_desktop\_session} signature).
This behavior has been noted in prior FormBook analyses~\cite{anyrun2023formbook}, where execution of \texttt{mstsc.exe} is considered a key indicator for hunting FormBook activity. However, under baseline sandbox execution, this behavior is not observable.

\begin{table*}[!htb]
\centering
\caption{Malware Family Detection: Existing Sandboxes vs LLM-based Config Extraction}
\label{tab:family-llm-comparison}
\footnotesize
\begin{threeparttable}
\begin{tabular}{l|rr|rrrrrrr|rrrrr|r}
\toprule[1.5pt]
 & \multicolumn{2}{c|}{\textbf{Baseline}} & \multicolumn{7}{c|}{\textbf{MalwareBazaar Collected Sandboxes}} & \multicolumn{5}{c|}{\textbf{\tool- single models vs ensemble models}} & \\
\textbf{Family} & \textbf{Joe} & \textbf{CAPE} & \textbf{ANY} & \textbf{MB-C} & \textbf{Int} & \textbf{Sig} & \textbf{Tri} & \textbf{VMR} & \textbf{RLab} & \textbf{DS} & \textbf{Gem} & \textbf{Lla} & \textbf{Qw} & \textbf{\tool} & \textbf{Best}$^\dagger$ \\
\midrule[0.8pt]
RedLine & 196 & 195 & 128 & 183 & 41 & 108 & 131 & 178 & 74 & 120 & 108 & 109 & 161 & 240 & +44 \\
Amadey & 140 & 140 & 86 & 1 & 134 & 107 & 77 & 115 & 13 & 119 & 101 & 92 & 134 & 198 & +58 \\
Formbook & 45 & 41 & 28 & 28 & 30 & 44 & 38 & 13 & 9 & 41 & 47 & 35 & 40 & 74 & +29 \\
StealC & 9 & 10 & 0 & 0 & 3 & 3 & 3 & 3 & 0 & 16 & 11 & 6 & 27 & 39 & +29 \\
Remcos & 10 & 10 & 3 & 7 & 5 & 10 & 9 & 3 & 5 & 7 & 8 & 6 & 13 & 23 & +13 \\
STOP & 7 & 7 & 2 & 3 & 2 & 6 & 1 & 2 & 1 & 5 & 8 & 2 & 9 & 19 & +12 \\
StormKitty & 1 & 1 & 0 & 0 & 0 & 1 & 1 & 0 & 0 & 3 & 4 & 0 & 7 & 13 & +12 \\
AsyncRAT & 1 & 1 & 0 & 0 & 0 & 0 & 0 & 1 & 1 & 3 & 4 & 0 & 7 & 13 & +12 \\
WorldWind & 0 & 1 & 0 & 0 & 0 & 0 & 0 & 0 & 0 & 3 & 4 & 0 & 7 & 13 & +12 \\
PYSA & 3 & 3 & 0 & 0 & 0 & 3 & 0 & 0 & 3 & 2 & 3 & 2 & 6 & 11 & +8 \\
SystemBC & 2 & 3 & 0 & 2 & 3 & 3 & 2 & 0 & 1 & 4 & 2 & 1 & 5 & 10 & +7 \\
Cuba & 1 & 1 & 0 & 0 & 0 & 1 & 0 & 0 & 1 & 4 & 3 & 0 & 3 & 8 & +7 \\
DLAgent05 & 0 & 4 & 0 & 0 & 0 & 0 & 0 & 0 & 0 & 4 & 5 & 4 & 4 & 8 & +4 \\
Vidar & 23 & 10 & 7 & 5 & 0 & 4 & 12 & 0 & 3 & 2 & 4 & 2 & 3 & 6 & -17 \\
Clop & 0 & 1 & 0 & 0 & 0 & 0 & 0 & 0 & 0 & 3 & 0 & 0 & 3 & 6 & +5 \\
\midrule[0.8pt]
\textbf{Total} & \textbf{484} & \textbf{457} & \textbf{266} & \textbf{248} & \textbf{240} & \textbf{332} & \textbf{309} & \textbf{329} & \textbf{331} & \textbf{353} & \textbf{331} & \textbf{274} & \textbf{446} & \textbf{712} & \textbf{+228} \\
\bottomrule[1.5pt]
\end{tabular}
\begin{tablenotes}
\footnotesize
\item \textbf{Baseline}: Joe = JoeSandbox, CAPE = CAPEv2 Baseline.
\item \textbf{MalwareBazaar}: ANY = ANY.RUN, MB-C = CAPE, Int = Intezer, Sig = MB-Signature, Tri = Triage, VMR = VMRay, RLab = ReversingLabs.
\item \textbf{\tool}: DS = DeepSeek-R1-7B Only, Gem = Gemma3-12B Only, Lla = LLama3.1-8B Only, Qw = Qwen3-8B Only, \tool = Ensemble configuration
\item $^\dagger$\textbf{Best}: \tool count minus best existing sandbox count. Top 15 families by LLM detection count shown. Sample counts per family.
\end{tablenotes}
\end{threeparttable}
\end{table*}

\subsubsection{Malware Family Characterization}
\tool identify new malware family classifications for 24 unique samples that are not reported by existing analysis platforms. As shown in Table~\ref{tab:family-llm-comparison} (complete list is shown in Appendix \ref{appendix:malfam}), we compare malware family coverage across publicly available reports from ANYRUN~\cite{anyrun2024}, CAPEv2 (both online and local deployments)~\cite{cape2024}, Joe Sandbox~\cite{joesandbox2024}, Intezer~\cite{intezer2024}, MalwareBazaar signatures~\cite{malwarebazaar_sample_caf}, Triage~\cite{triage2024}, VMRay~\cite{vmray2024}, and ReversingLabs~\cite{reversinglabs2024}.
For the top 15 malware families detected by \tool, our approach identifies more family matches than other sandbox, dynamic, or hybrid analysis solutions, achieving the highest coverage in 14 out of the 15 families. More broadly, when comparing total malware family coverage across all evaluated platforms with public reports, our framework identifies 47\% more malware families than the best-performing baseline (Joe Sandbox).

\begin{tcolorbox}[left=2pt,right=2pt,top=2pt,bottom=2pt,colback=gray!5,colframe=gray!50,boxrule=0.5pt,title={\small RQ3: Takeaway}]
\small
Across signature analysis, case studies, and malware family classification, \tool consistently expands behavioral visibility beyond baseline sandbox execution. Overall, these results show that automated evasion bypass not only increases the number of observable behaviors and identified malware families, but also improves the fidelity of malware characterization.
\end{tcolorbox}

\subsection{Performance Evaluation}
As described in Section~\ref{5.3}, \tool is evaluated on 334 malware samples through a total of 13,778 sandbox executions, with each execution subject to a maximum analysis time of 200 seconds. The total execution and signature-generation time for the ensemble configuration of \tool is $\approx 765 hours$. This corresponds to an effective analysis throughput of roughly ten (10) obfuscated malware samples per day, a scale that would typically require days or weeks of manual analysis per sample. We note that the total execution runtime is reported considering sequential execution of a single CAPEv2 instance on the same single host machine. In the case of parallelizing CAPEv2 instances and host machines, the overall analysis time can be reduced by one or more orders of magnitude, depending on available computational resources. Even without accounting for human error or analyst fatigue, the combination of a 79\% bypass success rate and substantially reduced analysis time demonstrates that \tool enables practical analysis of evasive malware at a scale that is infeasible with manual approaches.

\section{Limitation \& Future Work}
Our evaluation is limited to a dataset of Windows PE malware, including native binaries and .NET assemblies, which reflects mainstream sandbox environments but may limit generalization to other platforms or architectures. We rely on VirusTotal and CAPA annotations to identify sandbox-evasive samples; while widely used, these labels are not exhaustive and may miss novel dynamically triggered evasion techniques. We also observe that most generated bypass rules employ skip-based actions, likely because generating alternative control-flow manipulations is challenging for general-purpose LLMs under constrained context, which may explain the remaining 21\% of samples that do not successfully execute. Moreover, due to the absence of ground-truth datasets for complete evasion behavior or payload exposure, our evaluation emphasizes relative improvements over baseline sandbox execution rather than exhaustive behavioral discovery. Finally, although iterative refinement mitigates LLM generation errors, the framework inherits limitations of the underlying models, including occasional instability and increased computational cost during large-scale analysis.

For future work, we plan to adopt LLM fine-tuning to create specialized models using manually crafted YARA rules from open-source repos to enable the generation of rules with more complex and diverse actions. This enhancement is expected to improve overall framework effectiveness and reduce retry rates. Additionally, we plan to extend the framework to support other architectures and conduct broader evaluations across additional sandbox platforms and larger, more diverse malware datasets, including more complex samples.

\section{Related work}
\subsection{ Traditional Malware Analysis}  
Static malware analysis inspects code without executing it, but its effectiveness is often limited by obfuscation that alters or hides key features, as well as by anti-disassembly techniques that tricks disassemblers and corrupt disassembly output, resulting in uninterpreted byte streams \cite{sihwail2018survey, bakour2018android, aghakhani2020malware, yong2021inside, fereidooni2016anastasia, singh2018challenge,ali2015opseq}. For instance, Molina-Coronado et al.~\cite{molina2025light} show that obfuscation substantially distorts common APK-derived features in static analysis, while Chen et al.~\cite{chen2016advanced} report that targeted malware does not necessarily contain more anti-debugging or anti-VM evasion techniques compared to generic malware. Cameron et al. \cite{cameron2025statos} present StatOS, a USB-bootable environment that enables secure, user-friendly static malware analysis and sample acquisition. While static analysis is scalable, it often struggles to reveal true runtime behaviors, motivating dynamic analysis approaches that execute malware and observe system interactions. In practice, dynamic analysis executes samples in contained environments such as sandboxes and records traces such as API or system calls \cite{mahmoud2024redefining, cui2023droidhook, li2022dmalnet, yang2023goosebt}. However, many malware families often change runtime behavior when it detects sandbox, VM, or debugger, which makes authenticate feature extraction difficult \cite{yong2021inside, jeon2020dynamic, talukder2020survey, hussaini2021object, da2022exploring}. 
Previous work counters evasion using techniques such as dynamic binary instrumentation \cite{gaber2025defeating, park2019automatic}, hybrid instrumentation \cite{,ali2018toward,ali2016aspectdroid}, trace-based localization of evasive regions \cite{kirat2015malgene}, and sandbox-layer countermeasures to force to reveal more behavioral disclosure \cite{noor2018countering}. In contrast, our framework introduces an LLM-guided feedback loop that iteratively analyzes evasion behavior and refines detection logic using sandbox execution feedback. 
\subsection{Malware Analysis Using LLM }
Recent studies suggest that combining the AI-driven approach with traditional malware analysis can improve both accuracy and efficiency. In these analysis approaches, AI is commonly used to automate or assist routine steps—such as feature extraction, behavioral pattern analysis, and rule/signature generation to support faster threat detection and response \cite{djenna2023artificial, ibrahim2022method, saqib2024comprehensive, khatun2025androbyte, lea2025rex86, anderson2017evading, almomani2025behavioral, patil2021improving, walton2024exploring}. Building on this direction, LLMs are increasingly being used to interpret code artifacts and execution traces to support detection, classification, and analyst decision-making \cite{nam2024using, walton2024exploring, qian2025lamd, al2024exploring, feng2025llm, rondanini2025malware, he2025benchmarking, naseer2025obfuscated}.
For instance, Zhang et al.  \cite{zhang2025automatically} introduces an LLM-based tool that generates deployable YARA and Semgrep rules to detect malicious open-source software packages that addressed the limitations of manually written rules in supply-chain attacks.
Beyond rule generation, Patsakis et al. \cite{patsakis2024assessing} assess LLM-based malicious-code deobfuscation and report that LLMs can extract useful threat intelligence from heavily obfuscated PowerShell malware. In similar, Khalifa et al. \cite{khalifa2024sama} propose SAMA, which integrates static and sandbox analysis with ChatGPT-assisted YARA/Snort rule generation and family prediction. Compared with prior LLM-assisted approach, our framework focuses on evasion bypass by generating YARA rules using sandbox execution traces and refining them through validation plus sandbox feedback until addition malware behaviors are observed.

\section{Conclusion}
In this paper, we presented \tool, an automated, LLM-assisted feedback framework that actively bypasses sandbox evasion by iteratively generating, validating, and refining YARA rules grounded in execution feedback. Rather than relying on static heuristics or manual intervention, \tool operationalizes analyst-driven evasion bypass at scale.
Through empirical evaluation on 334 evasive malware samples across more than 13,000 sandbox executions, we show that the ensembled \tool substantially improves behavioral visibility, achieving a 79\% bypass success rate and uncovering 82 previously hidden malware behavioral signatures. Our ablation study demonstrates that execution-grounded iteration and model diversity are the primary drivers of effectiveness, while automated sanitization is essential for sustaining large-scale analysis. Finally, our qualitative analysis illustrate how \tool exposes behaviors and indicators that remain hidden under baseline CAPE sandbox analysis and enables the identification of 47\% more malware family classifications.

\section*{Ethical Considerations}

This work analyzes real-world malware and sandbox evasion techniques to improve defensive malware analysis. All experiments were conducted in strictly isolated, virtualized sandbox environments with controlled networking and without access to the internet to prevent unintended propagation or external impact. The malware samples were obtained from legitimate research repositories and used solely for defensive research purposes.

ABLE generates YARA rules that bypass sandbox evasion checks; these rules operate only within debugger-enabled analysis sandboxes and do not modify standard software or enable real-world exploitation. Their purpose is to reveal malware binary behaviors already present but hidden during baseline execution.\

ABLE relies on large language models (LLMs); therefore, we mitigate the risk of hallucinated or incorrect rule generation through deterministic sanitization, validation, and iterative sandbox feedback prior to rule deployment. Furthermore, \tool uses publicly available LLMs executed locally, which significantly reduces the risk, compared to online models, of producing or releasing live malware samples. Our research and framework are intended solely to support defensive and reproducible security research.

Finally, the authors are responsible for all content in this manuscript, including the ABLE framework design, experiments, and conclusions. AI-based tools (e.g., GPT/Claude and Grammarly) were used solely for language editing and proofreading and did not contribute technical ideas, analysis, or scientific claims. All literature review, citations, and experimental design were developed and verified independently by the authors.



\balance
\bibliographystyle{plainurl}
\bibliography{ref}

\cleardoublepage

\color{black}
\appendix
\section{Sandbox Evasion Evaluation}
\label{appendix:alkhaser}
\begin{table}[h]
\centering
\caption{Sandbox evasion test results using al-khaser.}
\label{tab:alkhaser}
\footnotesize
\setlength{\tabcolsep}{3pt}
\begin{tabular}{l|c|c|c|c}
\toprule
\textbf{Check} & \textbf{CAPE} & \textbf{AnyRun} & \textbf{Hybrid} & \textbf{QiAnXin} \\
\midrule
Hypervisor & $\times$ & & $\times$ & \\
Username & & $\times$ & & \\
\midrule
CacheMemory & $\times$ & $\times$ & & $\times$ \\
MemoryDevice & $\times$ & $\times$ & & $\times$ \\
Fan & $\times$ & $\times$ & $\times$ & $\times$ \\
TempProbe & $\times$ & $\times$ & $\times$ & $\times$ \\
VoltageProbe & $\times$ & $\times$ & $\times$ & $\times$ \\
PortConnector & $\times$ & $\times$ & & $\times$ \\
CIM\_Memory & $\times$ & $\times$ & & $\times$ \\
CIM\_Sensor & $\times$ & $\times$ & $\times$ & $\times$ \\
NumericSensor & $\times$ & $\times$ & $\times$ & $\times$ \\
TempSensor & $\times$ & $\times$ & $\times$ & $\times$ \\
VoltageSensor & $\times$ & $\times$ & $\times$ & $\times$ \\
PhysicalConn & $\times$ & $\times$ & & $\times$ \\
CIM\_Slot & $\times$ & $\times$ & & $\times$ \\
CompSystem VM & & & $\times$ & \\
CPU Count$<$2 & & $\times$ & & $\times$ \\
\midrule
\textbf{Failed} & \textbf{14} & \textbf{15} & \textbf{9} & \textbf{14} \\
\bottomrule
\end{tabular}
\end{table}

\section{Validation and Sanitization Algorithm}
\label{appendix:validation-algorithm}
\begin{algorithm}[H]
\caption{Validation and Sanitization Pipeline}
\label{alg:validation}
\begin{algorithmic}[1]
\Require $R$: generated rule, $i$: iteration index
\Ensure $\langle valid, R', E \rangle$: validity flag, sanitized rule, error set
\Function{Validate}{$R, i$}
    \State $R' \gets \Call{Sanitize}{R}$ \Comment{Apply auto-corrections}
    \State $E \gets \emptyset$
    \If{$\neg\Call{HasValidStructure}{R'}$}
        \State $E \gets E \cup \{\textsc{StructureError}\}$
    \EndIf
    \If{$\neg\Call{HasCapeOptions}{R'}$}
        \State $R' \gets \Call{InjectMetadata}{R'}$
    \EndIf
    \For{\textbf{each} pattern $p$ \textbf{in} $R'$}
        \If{$\Call{ContainsAssembly}{p}$}
            \State $E \gets E \cup \{\textsc{AssemblyConfusion}\}$
        \EndIf
        \If{$|p| < 6$ \textbf{or} $|p| > 20$}
            \State $E \gets E \cup \{\textsc{LengthError}\}$
        \EndIf
        \If{$\Call{WildcardRatio}{p} > 0.8$}
            \State $E \gets E \cup \{\textsc{SpecificityError}\}$
        \EndIf
        \If{$i = 0$ \textbf{and} $\Call{LacksContext}{p}$}
            \State $E \gets E \cup \{\textsc{ContextError}\}$
        \EndIf
    \EndFor
    \If{$\Call{HasDuplicates}{R'}$}
        \State $E \gets E \cup \{\textsc{DuplicateError}\}$
    \EndIf
    \State \Return $\langle E = \emptyset, R', E \rangle$
\EndFunction
\end{algorithmic}
\end{algorithm}

\section{\tool Sandbox Validation Algorithm}
\label{appendix:Sandbox-algorithm}
\begin{algorithm}[h]
\caption{Sandbox Behavioral Validation}
\label{alg:sandbox}
\begin{algorithmic}[1]
\Require YARA rule $r$, baseline signatures $B$, sample $s$
\Ensure Validation result $\langle status, data \rangle$
\Function{ValidateInSandbox}{$r, B, s$}
    \State $\Call{DeployRule}{sandbox, r}$
    \State $result \gets \Call{Execute}{sandbox, s}$
    \State $hit \gets \Call{CheckYaraMatch}{result}$
    \State $S_{new} \gets result.signatures $
    \State $S_{crash} \gets \Call{GetCrashSignatures}{result}$
    \If{$S_{crash} \neq \emptyset$}
        \State \Return $\langle \textsc{Crashed}, result.debugger\_log \rangle$
    \ElsIf{$hit \land |S_{new}| \geq \theta$}
        \State \Return $\langle \textsc{Success}, S_{new} \rangle$
    \ElsIf{$hit$}
        \State \Return $\langle \textsc{Partial}, \text{``No new behaviors''} \rangle$
    \Else
        \State \Return $\langle \textsc{Failed}, \text{``Pattern not matched''} \rangle$
    \EndIf
\EndFunction
\end{algorithmic}
\end{algorithm}

\section{\tool Feedback Iteration Algorithm}
\label{appendix:feedback-algorithm}
\begin{algorithm}[H]
\caption{Feedback-Driven Rule Iteration}
\label{alg:evolution}
\begin{algorithmic}[1]
\Require Trace $T$, baseline $B$, max iterations $N$
\Ensure Best bypass rule or failure
\Function{ReviseRule}{$T, B, N$}
    \State $hist \gets []$
    \State $best \gets \textsc{None}$
    \For{$i \gets 1$ \textbf{to} $N$}
        \If{$i = 1$}
            \State $prompt \gets \Call{FormatInitialPrompt}{T}$
        \Else
            \State $feedback \gets \Call{AnalyzeFailure}{}$
            \State $prompt \gets \Call{BuildPrompt}{T, hist, best, feedback}$
        \EndIf
        \State $rule \gets \Call{LLM.Generate}{prompt}$
        \State $rule \gets \Call{ValidateAndSanitize}{rule}$
        \State $result \gets \Call{ValidateInSandbox}{rule, B}$
        \State $hist.\text{append}(\langle rule, result \rangle)$
        \If{$result.status = \textsc{Success}$}
            \State \Return $rule$
        \ElsIf{$result.signatures > best.signatures$}
            \State $best \gets rule$
        \EndIf
    \EndFor
    \State \Return $best$ \textbf{or} \textsc{Failure}
\EndFunction
\end{algorithmic}
\end{algorithm}
\newpage
\onecolumn
\section{\tool Workflow Example for StealC Malware}
\label{appendix:workflowexample}
\begin{figure*}[ht]
\centering
\resizebox{\textwidth}{!}{%
\begin{tikzpicture}[
    node distance=0.25cm and 0.6cm,
    tracebox/.style={rectangle, draw=black!60, fill=gray!3, rounded corners=2pt, text width=4.5cm, align=left, inner sep=4pt, font=\ttfamily\tiny},
    rulebox/.style={rectangle, draw=blue!60, fill=blue!5, rounded corners=2pt, text width=4.5cm, align=left, inner sep=4pt, font=\ttfamily\tiny},
    iterbox/.style={rectangle, draw=teal!60, fill=teal!5, rounded corners=2pt, text width=4.5cm, align=left, inner sep=4pt, font=\small},
    successiterbox/.style={rectangle, draw=green!60!black, fill=green!8, rounded corners=2pt, text width=4.5cm, align=left, inner sep=4pt, font=\small},
    failiterbox/.style={rectangle, draw=red!60, fill=red!5, rounded corners=2pt, text width=4.5cm, align=left, inner sep=4pt, font=\small},
    feedbackbox/.style={rectangle, draw=orange!60, fill=orange!5, rounded corners=2pt, text width=4.5cm, align=left, inner sep=4pt, font=\small},
    logbox/.style={rectangle, draw=gray!50, fill=gray!3, rounded corners=2pt, text width=4.5cm, align=left, inner sep=3pt, font=\ttfamily\tiny},
    stagelabel/.style={font=\bfseries\small, fill=white, inner sep=3pt},
    arrow/.style={-{Stealth[length=2.5mm]}, thick, draw=black!60},
    successarrow/.style={-{Stealth[length=2.5mm]}, thick, draw=green!60!black},
    analyzearrow/.style={-{Stealth[length=2.5mm]}, thick, draw=orange!70},
    evolvearrow/.style={-{Stealth[length=2.5mm]}, thick, draw=teal!70},
]
\node[stagelabel, fill=violet!10, draw=violet!50, rounded corners=2pt] (label1) {Stage 1: Trace Analysis + Rule Generation};
\node[tracebox, below=0.2cm of label1] (trace) {
    \textcolor{gray}{// SHA256: 77d6f1914af6caf9...}\\
    \textcolor{gray}{// CAPE Debugger Trace - ExitProcess@0010CF7E}\\[1pt]
    \textcolor{gray!70}{0010CF89  55           PUSH EBP}\\
    \textcolor{gray!70}{0010CF8A  8BEC         MOV EBP, ESP}\\
    0010CF97  E87349FFFF   CALL 0x10190F\\
    0010CF9C  7403         JZ 0x5\\
    0010CFE5  E8A9010000   CALL 0x10D193\\
    0010D07A  FF15DC3531   CALL OpenEventA\\
    0010D080  3BC7         CMP EAX, EDI\\
    0010D082  75E1         JNZ 0xFFFFFFE3\\
    0010D088  FF15243731   CALL CreateEventA\\
    0010D090  7403         JZ 0x5\\
    0010D095  E830FEFFFF   CALL 0x10CECA\\
    \textcolor{gray!60}{--- Time Check Function ---}\\
    0010CF0E  FF15DC3631   CALL GetSystemTime\\
    0010CF32  FF30         PUSH [EAX] \textcolor{gray}{"04/02/2023"}\\
    0010CF34  FF155C3731   CALL sscanf\\
    0010CF50  FF15E43631   CALL SysTimeToFileTime\\
    0010CF6A  3B442414     CMP EAX, [ESP+0x14]\\
    0010CF6E  7214         JB 0x16\\
    0010CF7C  6A00         PUSH 0x0\\
    00 10 CF 7E FF 15 BC 36 31   CALL ExitProcess
};
\node[rulebox, below=0.15cm of trace, draw=violet!60, fill=violet!5] (prompt0) {
    \textbf{V2 Prompt: Counterfactual Analysis}\\[2pt]
    \textbf{Phase 1:} Establish factual outcome\\
    {\scriptsize $\rightarrow$ ``The program exits at ExitProcess''}\\
    \textbf{Phase 2:} ``What if instruction X skipped?''\\
    {\scriptsize $\rightarrow$ Hypothesize each CMP/JB effect}\\
    \textbf{Phase 3:} Necessity + Sufficiency test\\
    {\scriptsize $\rightarrow$ Is CMP necessary for exit?}\\
    \textbf{Phase 4:} Select 3 bypass candidates\\
    {\scriptsize $\rightarrow$ Rank by causal impact}\\
    \textbf{Phase 5:} Build patterns (6-20 bytes)\\
    {\scriptsize $\rightarrow$ Include context for uniqueness}\\[1pt]
};
\node[rulebox, below=0.15cm of prompt0, draw=gray!60, fill=gray!5, text width=4.5cm] (constraints) {
    \textbf{Output Constraints:}\\[2pt]
    {\scriptsize \textbf{Length:} 6--20 bytes (2--3 instructions)}\\
    {\scriptsize \textbf{Wildcards:} \texttt{??} for addr/offset only}\\
    {\scriptsize \textbf{Source:} From actual trace data}\\
    {\scriptsize \textbf{Metadata:} \texttt{cape\_options} required}\\[2pt]
    \textcolor{gray!70}{\textbf{Format:}}\\
    {\ttfamily\tiny rule Bypass\_Sample \{}\\
    {\ttfamily\tiny ~~meta: cape\_options = "bp0=..."}\\
    {\ttfamily\tiny ~~strings: \$p0 = \{E8 ?? ?? ...\}}\\
    {\ttfamily\tiny ~~condition: any of them \}}
};
\node[failiterbox, below=0.15cm of constraints, text width=4.5cm] (baseline) {
    \textbf{Baseline Signatures:}\\[1pt]
    \textbf{Total:} 7 signatures\\[1pt]
    {\scriptsize \textbullet~antivm\_checks\_available\_memory}\\
    {\scriptsize \textbullet~stealth\_timeout}\\
    {\scriptsize \textbullet~queries\_locale\_api}\\
    {\scriptsize \textbullet~queries\_computer\_name}\\
    {\scriptsize \textbullet~language\_check\_registry}\\
    {\scriptsize \textbullet~queries\_user\_name}\\
    {\scriptsize \textbullet~injection\_rwx}\\[1pt]
    \textbf{Malware Families:} None\\
};
\node[stagelabel, fill=teal!10, draw=teal!50, rounded corners=2pt, right=0.8cm of label1] (label2) {Stage 2: Sandbox Validation (Iteration Loop)};
\node[failiterbox, below=0.2cm of label2, text width=5.2cm] (iter0) {
    \textbf{Iteration 0} \hfill \textcolor{red!70}{$\times$ FAILED}\\[2pt]
    {\ttfamily\tiny rule Bypass\_Sample \{}\\
    {\ttfamily\tiny ~~meta:\\ cape\_options = "bp0=\$pattern0+0,action0=skip,bp1=\$pattern1+0,action1=skip,\\ bp2=\$pattern2+0,action2=skip,count=0"}\\
    {\ttfamily\tiny ~~strings:}\\
    {\ttfamily\tiny ~~~~\$p0 = \{6A 00 FF 15 ?? ?? ?? ??\}}\\
    {\ttfamily\tiny ~~~~\$p1 = \{3B C7 75 E1 ?? ?? ?? ??\}}\\
    {\ttfamily\tiny ~~~~\$p2 = \{53 57 57 57 FF 15 ?? ?? ?? ??\}}\\
    {\ttfamily\tiny ~~condition: any of them}\\
    {\ttfamily\tiny \}}\\[1pt]
    {\ttfamily\tiny \textbf{Result:} 7 $\rightarrow$ 7 sigs | \textbf{Family:} None}
};
\node[failiterbox, below=0.15cm of iter0, text width=5.2cm] (iter1) {
    \textbf{Iteration 1} \hfill \textcolor{red!70}{$\times$ FAILED}\\[2pt]
    {\ttfamily\tiny rule Bypass\_Sample \{}\\
    {\ttfamily\tiny ~~meta:\\ cape\_options = "bp0=\$pattern0+0,action0=skip,bp1=\$pattern1+0,action1=skip,\\ bp2=\$pattern2+0,action2=skip,count=0"}\\
    {\ttfamily\tiny ~~strings:}\\
    {\ttfamily\tiny ~~~~\$p0 = \{FF 15 ?? ?? ?? ?? 3B C7 75 E1\}}\\
    {\ttfamily\tiny ~~~~\$p1 = \{FF 15 ?? ?? ?? ?? 8B F0 74 03\}}\\
    {\ttfamily\tiny ~~~~\$p2 = \{74 03 75 01 FF 15 ?? ?? ?? ??\}}\\
    {\ttfamily\tiny ~~condition: any of them}\\
    {\ttfamily\tiny \}}\\[1pt]
    {\ttfamily\tiny \textbf{Result:} 7 $\rightarrow$ 7 sigs | \textbf{Family:} None}
};
\node[successiterbox, below=0.15cm of iter1, text width=5.2cm] (iter2) {
    \textbf{Iteration 2} \hfill \textcolor{green!60!black}{\checkmark SUCCESS}\\[2pt]
    {\ttfamily\tiny rule Bypass\_Sample \{}\\
    {\ttfamily\tiny ~~meta:\\ cape\_options = "bp0=\$pattern0+0,action0=skip,bp1=\$pattern1+0,action1=skip,\\ bp2=\$pattern2+0,action2=skip,count=0"}\\
    {\ttfamily\tiny ~~strings:}\\
    {\ttfamily\tiny ~~~~\$p0 = \{E8 ?? ?? ?? ?? 74 03\}}\\
    {\ttfamily\tiny ~~~~\$p1 = \{E8 ?? ?? ?? ?? 75 01\}}\\
    {\ttfamily\tiny ~~~~\$p2 = \{E8 A9 01 00 00 50 74 03\}}\\
    {\ttfamily\tiny ~~condition: any of them}\\
    {\ttfamily\tiny \}}\\[1pt]
    {\ttfamily\tiny \textbf{Result:} 7 $\rightarrow$ 11 sigs (+4) | \textbf{Family:} StealC}
};
\node[stagelabel, fill=orange!10, draw=orange!50, rounded corners=2pt, right=0.8cm of label2] (label3) { Stage 3: LLM Analysis And Feedback Evolve};
\node[failiterbox, below=0.2cm of label3, text width=4.8cm] (fail0) {
    \textbf{Iteration 0 Analysis:}\\[2pt]
    \textbf{failure\_reason:}\\
    {\scriptsize ``Rule hit but no new signatures''}
};
\node[logbox, below=0.1cm of fail0, text width=4.8cm] (log0) {
    \textbf{debugger\_log:}\\
    Breakpoint 1 hit at 0x00F4D080\\
    0x00F4D080 3B C7 CMP EAX, EDI\\
    ActionDispatcher: skipping instruction.\\
    Breakpoint 2 hit at 0x00F4D084\\
    0x00F4D084 53 PUSH EBX\\
    ActionDispatcher: skipping instruction.
};
\node[feedbackbox, below=0.1cm of log0, text width=4.8cm] (suggest0) {
    \textbf{improvement\_suggestions:}\\[1pt]
    {\scriptsize ``Refine patterns to target specific}\\
    {\scriptsize anti-analysis checks (e.g., ExitProcess)}\\
    {\scriptsize and adjust skip offsets to avoid}\\
    {\scriptsize invalid code jumps.''}
};
\node[failiterbox, below=0.15cm of suggest0, text width=4.8cm] (fail1) {
    \textbf{Iteration 1 Analysis:}\\[2pt]
    \textbf{failure\_reason:}\\
    {\scriptsize ``Rule hit but no new signatures''}
};
\node[logbox, below=0.1cm of fail1, text width=4.8cm] (log1) {
    \textbf{debugger\_log:}\\
    Breakpoint 0 hit at 0x001AD07A\\
    0x001AD07A FF 15 DC 35 3B 00 CALL [IAT]\\
    ActionDispatcher: skipping instruction.\\
    Breakpoint 1 hit at 0x001AD088\\
    0x001AD088 FF 15 24 37 3B 00 CALL [IAT]\\
    ActionDispatcher: skipping instruction.
};
\node[feedbackbox, below=0.1cm of log1, text width=4.8cm] (suggest1) {
    \textbf{improvement\_suggestions:}\\[1pt]
    {\scriptsize ``Adjust skip offset to avoid invalid jumps.}\\
    {\scriptsize Target relative calls (E8) instead of}\\
    {\scriptsize absolute calls (FF 15) to align}\\
    {\scriptsize with actual code flow.''}
};
\node[successiterbox, below=0.15cm of suggest1, text width=4.8cm] (success2) {
    \textbf{Iteration 2:} \textcolor{green!60!black}{\checkmark SUCCESS}\\[2pt]
    \textbf{New Behaviors Discovered (+4):}\\[1pt]
    {\scriptsize \textbullet~enumerates\_running\_processes}\\
    {\scriptsize \textbullet~http\_request \textcolor{red!70}{(C2: 162.0.238.10)}}\\
    {\scriptsize \textbullet~dead\_connect}\\
    {\scriptsize \textbullet~recon\_programs}\\[1pt]
    \textbf{Malware Family:} \textcolor{red!70}{StealC}
};
\draw[arrow] (baseline.east) -- ++(0.25,0) |- (iter0.west);
\draw[analyzearrow] ([yshift=-3pt]iter0.east) -- ++(0.55,0) |- ([yshift=3pt]fail0.west) node[pos=0.75, above, font=\tiny\sffamily, text=orange!70] {analyze};
\draw[analyzearrow] ([yshift=-3pt]iter1.east) -- ++(0.55,0) |- ([yshift=3pt]fail1.west) node[pos=0.75, above, font=\tiny\sffamily, text=orange!70] {analyze};
\draw[successarrow] ([yshift=-4.6pt]iter2.east) -- ++(0.55,0) |- ([yshift=4pt]success2.west) node[pos=0.75, above, font=\tiny\sffamily, text=green!60!black] {success};
\draw[evolvearrow] ([yshift=-3pt]suggest0.west) -- ++(-0.55,0) |- ([yshift=40pt]iter1.east) node[pos=0.25, below, font=\tiny\sffamily, text=teal!70] {evolve};
\draw[evolvearrow] ([yshift=-3pt]suggest1.west) -- ++(-0.55,0) |- ([yshift=15pt]iter2.east) node[pos=0.25, below, font=\tiny\sffamily, text=teal!70] {evolve};
\end{tikzpicture}
}%
\caption{This figures illustrate the workflow example of \tool on a StealC malware. \textbf{Stage 1:} Baseline execution detects only 7 generic signatures without family attribution; the LLM analyzes the trace and identifies time-based evasion logic. \textbf{Stage 2:} Iteration 0 attempts to bypass variable comparisons but fails. Iteration 1 targets absolute calls (\texttt{FF 15}) near the exit point; the debugger log shows breakpoints hit but evasion remains intact. \textbf{Stage 3:} The LLM suggests targeting relative calls (\texttt{E8}) instead, hypothesizing these invoke internal evasion-checking functions. Iteration 2 adopts this strategy, revealing 4 hidden behaviors including C2 communication (162.0.238.10) and correctly identifying the sample as \textbf{StealC}.}
\label{fig:case_study}
\end{figure*}
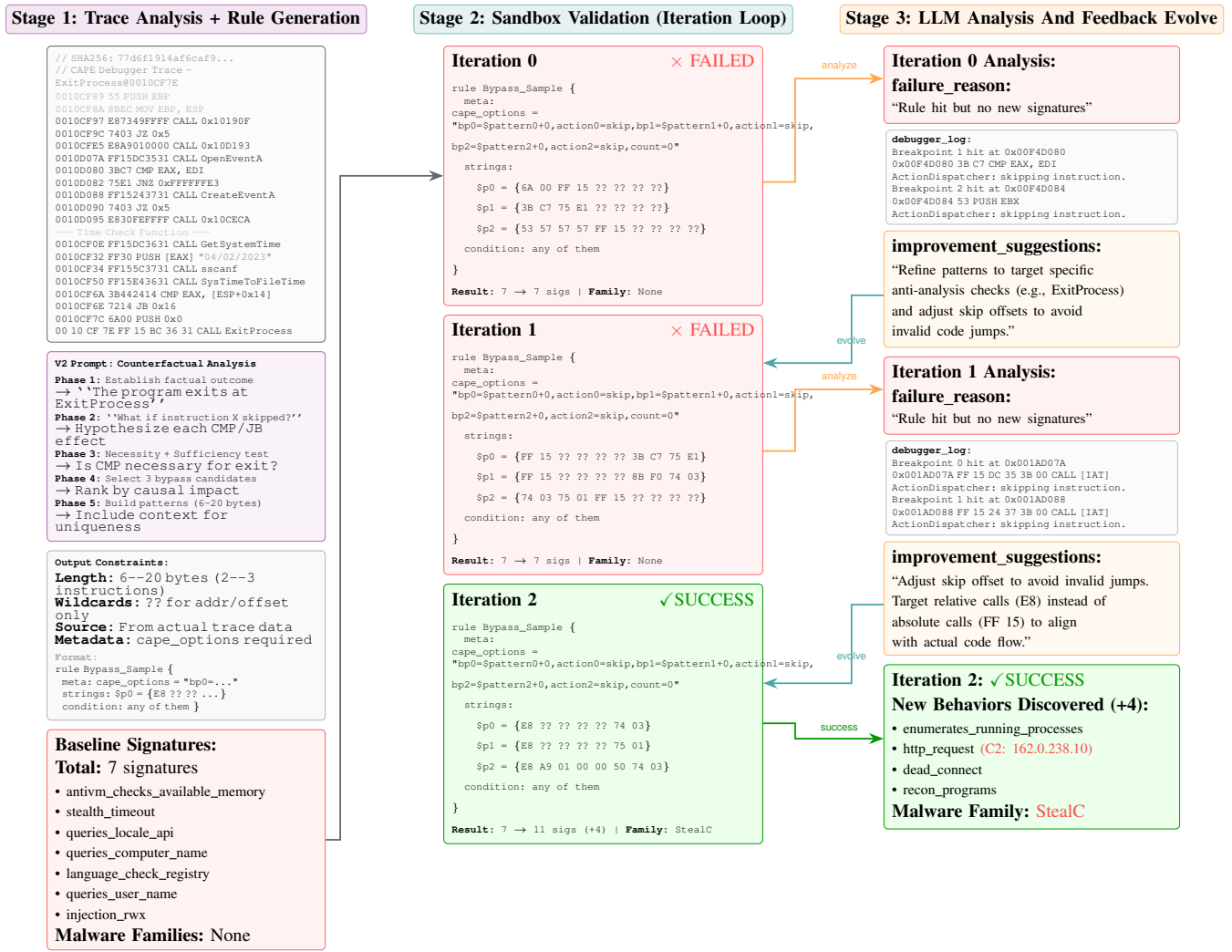

\newpage

\section{Complete List of Hidden Signatures Uncovered by \tool}
\label{appendix:newsig}
\begin{longtable}{@{}c@{\hspace{3pt}}l@{\hspace{1pt}}p{6cm}@{\hspace{6pt}}r@{\hspace{6pt}}r@{}}
\caption{Complete list of new signatures triggered during bypass success, grouped by CAPEv2 severity.
Severity scale: 1=Informational, 2=Low, 3=Medium.}
\small
\label{tab:signatures-actual-longtable} \\
\toprule
\textbf{Sev.} & \textbf{Signature} & \textbf{Description} & \textbf{Occ.} & \textbf{Samp.} \\
\midrule
\endfirsthead

\multicolumn{5}{c}{\tablename\ \thetable{}} \\
\toprule
\textbf{Sev.} & \textbf{Signature} & \textbf{Description} & \textbf{Occ.} & \textbf{Samp.} \\
\midrule
\endhead

\midrule
\multicolumn{5}{r}{Continued on the next page ...} \\
\endfoot

\midrule
\multicolumn{2}{l}{\textbf{Total: 82 signatures}} & & \textbf{3375} & \\
\bottomrule
\endlastfoot

\multicolumn{5}{l}{\textbf{Severity 3 (Medium) - 40 signatures}} \\
\midrule
3 & \texttt{antisandbox\_unhook} & Tries to unhook or modify Windows functions monitored by CAPE & 172 & 65 \\
3 & \texttt{antivm\_generic\_system} & Checks the system manufacturer, likely for anti-virtualization & 109 & 54 \\
3 & \texttt{enumerates\_physical\_drives} & Enumerates physical drives & 93 & 40 \\
3 & \texttt{physical\_drive\_access} & Attempted to write directly to a physical drive & 93 & 40 \\
3 & \texttt{procmem\_yara} & Yara detections observed in process dumps, payloads or dropped files & 81 & 24 \\
3 & \texttt{suspicious\_command\_tools} & Uses suspicious command line tools or Windows utilities & 58 & 33 \\
3 & \texttt{infostealer\_cookies} & Touches a file containing cookies, possibly for information gathering & 54 & 35 \\
3 & \texttt{persistence\_autorun\_tasks} & Installs itself for autorun at Windows startup & 35 & 25 \\
3 & \texttt{antiav\_detectfile} & Attempts to identify installed AV products by installation directory & 35 & 25 \\
3 & \texttt{cmdline\_obfuscation} & Appears to use command line obfuscation & 35 & 25 \\
3 & \texttt{recon\_beacon} & A process sent information about the computer to a remote location. & 31 & 24 \\
3 & \texttt{infostealer\_browser} & Steals private information from local Internet browsers & 27 & 9 \\
3 & \texttt{recon\_fingerprint} & Collects information to fingerprint the system & 16 & 6 \\
3 & \texttt{registry\_credential\_store\_access} & Accessed credential storage registry keys & 16 & 8 \\
3 & \texttt{antiav\_servicestop} & Attempts to stop active services & 15 & 6 \\
3 & \texttt{disables\_windowsupdate} & Attempts to disable Windows Auto Updates & 14 & 5 \\
3 & \texttt{deletes\_executed\_files} & Deletes executed files from disk & 13 & 3 \\
3 & \texttt{injection\_write\_exe\_process} & Writes an executable to the memory of another process & 11 & 3 \\
3 & \texttt{static\_pe\_anomaly} & Anomalous binary characteristics & 10 & 3 \\
3 & \texttt{multiple\_useragents} & Network activity contains more than one unique useragent. & 8 & 3 \\
3 & \texttt{removes\_zoneid\_ads} & Attempts to remove evidence of file being downloaded from the Internet & 8 & 2 \\
3 & \texttt{persistence\_autorun} & Installs itself for autorun at Windows startup & 8 & 3 \\
3 & \texttt{pe\_compile\_timestomping} & Binary compilation timestomping detected & 7 & 2 \\
3 & \texttt{binary\_yara} & Binary file triggered multiple YARA rules & 7 & 2 \\
3 & \texttt{network\_fake\_useragent} & Fake User-Agent detected & 6 & 1 \\
3 & \texttt{infostealer\_keylog} & Sniffs keystrokes & 6 & 5 \\
3 & \texttt{cape\_extracted\_content} & CAPE detected injection into a browser process, likely for Man-In-Browser (MITB) infostealing & 5 & 2 \\
3 & \texttt{recon\_programs} & Collects information about installed applications & 5 & 2 \\
3 & \texttt{virus} & Likely virus infection of existing binary & 4 & 1 \\
3 & \texttt{antisandbox\_mouse\_hook} & Installs an hook procedure to monitor for mouse events & 4 & 3 \\
3 & \texttt{warzonerat\_files} & Accesses or creates Warzone RAT directories and/or files & 4 & 1 \\
3 & \texttt{persistence\_ads} & Attempts to interact with an Alternate Data Stream (ADS) & 3 & 1 \\
3 & \texttt{infostealer\_ftp} & Harvests credentials from local FTP client softwares & 3 & 1 \\
3 & \texttt{injection\_runpe} & Executed a process and injected code into it, probably while unpacking & 3 & 1 \\
3 & \texttt{uses\_remote\_desktop\_session} & Connects to/from or queries a remote desktop session & 1 & 1 \\
3 & \texttt{deletes\_self} & Deletes its original binary from disk & 1 & 1 \\
3 & \texttt{antidebug\_windows} & Checks for the presence of known windows from debuggers and forensic tools & 1 & 1 \\
3 & \texttt{antiav\_avast\_libs} & Detects Avast Antivirus through the presence of a library & 1 & 1 \\
3 & \texttt{antisandbox\_sboxie\_libs} & Detects Sandboxie through the presence of a library & 1 & 1 \\
3 & \texttt{antivm\_generic\_diskreg} & Checks the presence of disk drives in the registry, possibly for anti-virtualization & 1 & 1 \\
\midrule
\multicolumn{5}{l}{\textbf{Severity 2 (Low) - 31 signatures}} \\
\midrule
2 & \texttt{terminates\_remote\_process} & Terminates another process & 414 & 105 \\
2 & \texttt{reads\_self} & Reads data out of its own binary image & 129 & 80 \\
2 & \texttt{reads\_memory\_remote\_process} & Reads from the memory of another process & 75 & 31 \\
2 & \texttt{injection\_rwx} & Creates RWX memory & 73 & 20 \\
2 & \texttt{packer\_entropy} & The binary likely contains encrypted or compressed data & 72 & 21 \\
2 & \texttt{process\_interest} & Expresses interest in specific running processes & 67 & 32 \\
2 & \texttt{enumerates\_running\_processes} & Enumerates running processes & 63 & 29 \\
2 & \texttt{uses\_windows\_utilities} & Uses Windows utilities for basic functionality & 58 & 33 \\
2 & \texttt{dynamic\_function\_loading} & Dynamic (imported) function loading detected & 55 & 31 \\
2 & \texttt{resumethread\_remote\_process} & Resumed a thread in another process & 50 & 30 \\
2 & \texttt{stealth\_window} & A process created a hidden window & 49 & 30 \\
2 & \texttt{creates\_suspended\_process} & Creates a process in a suspended state, likely for injection & 48 & 24 \\
2 & \texttt{http\_request} & Performs HTTP requests potentially not found in PCAP. & 47 & 29 \\
2 & \texttt{injection\_write\_process} & Writes to the memory another process & 45 & 27 \\
2 & \texttt{antisandbox\_sleep} & A process attempted to delay the analysis task. & 42 & 28 \\
2 & \texttt{dropper} & Drops a binary and executes it & 40 & 12 \\
2 & \texttt{uses\_windows\_utilities\_to\_create\_scheduled\_task} & Uses Windows utilities to create a scheduled task & 34 & 25 \\
2 & \texttt{mouse\_movement\_detect} & Checks for mouse movement & 32 & 17 \\
2 & \texttt{createtoolhelp32snapshot\_module\_enumeration} & Enumerates the modules from a process (may be used to locate base addresses in process injection) & 28 & 18 \\
2 & \texttt{dead\_connect} & Attempts to connect to a dead IP:Port (1 unique times) & 27 & 9 \\
2 & \texttt{antidebug\_guardpages} & Guard pages use detected - possible anti-debugging. & 17 & 6 \\
2 & \texttt{anomalous\_deletefile} & Anomalous file deletion behavior detected (10+) & 14 & 10 \\
2 & \texttt{suspicious\_communication\_trusted\_site} & Suspicious communication with abused trusted site & 10 & 3 \\
2 & \texttt{accesses\_recyclebin} & Manipulates data from or to the Recycle Bin & 9 & 2 \\
2 & \texttt{process\_needed} & Repeatedly searches for a not-found process, may want to run with startbrowser=1 option & 8 & 2 \\
2 & \texttt{network\_connection\_via\_suspicious\_process} & Attempts to make a network connection via suspicious process & 7 & 1 \\
2 & \texttt{contains\_pe\_overlay} & The PE file contains an overlay & 6 & 2 \\
2 & \texttt{credential\_access\_via\_windows\_credential\_history} & Attempts to access Users Windows Credential History File that is used by Microsoft's DPAPI & 5 & 1 \\
2 & \texttt{encrypted\_ioc} & At least one IP Address, Domain, or File Name was found in a crypto call & 4 & 2 \\
2 & \texttt{packer\_unknown\_pe\_section\_name} & The binary contains an unknown PE section name indicative of packing & 4 & 1 \\
2 & \texttt{network\_anomaly} & Network anomalies occured during the analysis. & 1 & 1 \\
\midrule
\multicolumn{5}{l}{\textbf{Severity 1 (Informational) - 11 signatures}} \\
\midrule
1 & \texttt{queries\_keyboard\_layout} & Queries the keyboard layout & 261 & 109 \\
1 & \texttt{queries\_locale\_api} & Queries the computer locale & 156 & 83 \\
1 & \texttt{stealth\_timeout} & Possible date expiration check, exits too soon after checking local time & 143 & 73 \\
1 & \texttt{queries\_computer\_name} & Queries computer hostname & 73 & 36 \\
1 & \texttt{static\_pe\_pdbpath} & The PE file contains a PDB path & 69 & 20 \\
1 & \texttt{antivm\_checks\_available\_memory} & Checks available memory & 65 & 26 \\
1 & \texttt{queries\_user\_name} & Queries the username & 22 & 9 \\
1 & \texttt{antidebug\_setunhandledexceptionfilter} & SetUnhandledExceptionFilter detected & 16 & 5 \\
1 & \texttt{accesses\_public\_folder} & A file was accessed within the Public folder. & 15 & 6 \\
1 & \texttt{language\_check\_registry} & Checks system language via registry key & 12 & 4 \\
1 & \texttt{cmdline\_terminate} & Executed a command line with /C or /R argument to terminate command shell on completion which can be used to hide execution & 5 & 3\\
\end{longtable}

\section{Top Malware Signatures Across LLMs Uncovered by \tool}
\label{appendix:newsig2}

\begin{table*}[!htb]
\caption{Top 15 New Signatures by Model and Prompt Strategy}
\centering
\scriptsize

\begin{threeparttable}
\begin{tabular}{l|rrrr|rrrr|rrrr|rrrr|r}
\toprule[1.5pt]
 & \multicolumn{4}{c|}{Qwen3-8B} & \multicolumn{4}{c|}{Llama3.1-8B} & \multicolumn{4}{c|}{Gemma3-12B} & \multicolumn{4}{c|}{DeepSeek-R1-7B} &  \\
\textbf{Signature} & \textbf{v0} & \textbf{v1} & \textbf{v2} & \textbf{v3} & \textbf{v0} & \textbf{v1} & \textbf{v2} & \textbf{v3} & \textbf{v0} & \textbf{v1} & \textbf{v2} & \textbf{v3} & \textbf{v0} & \textbf{v1} & \textbf{v2} & \textbf{v3} & \textbf{Tot} \\
\midrule[0.8pt]
terminates\_remote\_process & 21 & 20 & 21 & 24 & 19 & 18 & 21 & 15 & 32 & 44 & 46 & 36 & 27 & 20 & 22 & 28 & 414 \\
queries\_keyboard\_layout & 38 & 33 & 28 & 24 & 20 & 19 & 13 & 14 & 11 & 8 & 11 & 11 & 9 & 7 & 8 & 7 & 261 \\
antisandbox\_unhook & 7 & 9 & 8 & 9 & 11 & 4 & 8 & 5 & 22 & 26 & 24 & 20 & 5 & 2 & 4 & 8 & 172 \\
queries\_locale\_api & 28 & 29 & 24 & 18 & 9 & 6 & 5 & 7 & 4 & 2 & 3 & 3 & 6 & 4 & 5 & 3 & 156 \\
stealth\_timeout & 9 & 14 & 12 & 12 & 14 & 6 & 9 & 7 & 9 & 11 & 10 & 10 & 5 & 4 & 6 & 5 & 143 \\
reads\_self & 17 & 15 & 17 & 15 & 6 & 2 & 3 & 8 & 4 & 4 & 3 & 1 & 6 & 8 & 12 & 8 & 129 \\
antivm\_generic\_system & 7 & 9 & 9 & 10 & 6 & 3 & 4 & 3 & 7 & 12 & 8 & 6 & 5 & 7 & 5 & 8 & 109 \\
enumerates\_physical\_drives & 4 & 5 & 4 & 4 & - & 1 & - & 2 & 16 & 15 & 14 & 12 & 7 & 4 & 1 & 4 & 93 \\
physical\_drive\_access & 4 & 5 & 4 & 4 & - & 1 & - & 2 & 16 & 15 & 14 & 12 & 7 & 4 & 1 & 4 & 93 \\
procmem\_yara & 12 & 8 & 12 & 10 & 3 & 4 & 4 & 4 & 1 & 2 & 2 & 1 & 6 & 3 & 4 & 5 & 81 \\
reads\_memory\_remote\_process & 2 & 4 & 3 & 2 & 3 & 2 & 4 & 3 & 8 & 9 & 7 & 8 & 1 & 7 & 4 & 8 & 75 \\
queries\_computer\_name & 6 & 6 & 5 & 4 & 6 & 3 & 3 & 2 & 6 & 6 & 7 & 6 & 2 & 4 & 4 & 3 & 73 \\
injection\_rwx & 10 & 8 & 10 & 10 & 3 & 2 & 2 & 2 & 1 & 1 & 1 & 1 & 6 & 5 & 6 & 5 & 73 \\
packer\_entropy & 9 & 11 & 11 & 11 & 1 & 2 & 1 & 1 & - & - & - & - & 4 & 6 & 8 & 7 & 72 \\
static\_pe\_pdbpath & 10 & 12 & 12 & 12 & - & 1 & 1 & 1 & - & - & - & - & 3 & 4 & 7 & 6 & 69 \\
\midrule[0.8pt]
\textbf{Total} & 184 & 188 & 180 & 169 & 101 & 74 & 78 & 76 & 137 & 155 & 150 & 127 & 99 & 89 & 97 & 109 & 2013 \\
\bottomrule[1.5pt]
\end{tabular}
\begin{tablenotes}
\item
Each cell shows successful bypasses triggering the signature for that model/prompt combination.
\item
\textbf{v0-v3}: Prompt strategy versions. \textbf{Tot}: Total across all models and prompts.
\end{tablenotes}

\end{threeparttable}
\label{tab:signature-by-prompt}
\end{table*}

\section{Hidden Malware Signatures Distribution Uncovered by \tool}
\label{appendix:newsig3}
\begin{table*}[!htb]
\caption{Signature Categories and Distribution Across Models}
\centering
\footnotesize

\begin{threeparttable}
\begin{tabular}{l|rrrr|r|r}
\toprule[1.5pt]
 & \multicolumn{4}{c|}{\textbf{Count by Model}} & & \\
\textbf{Category} & \textbf{Qwen3-8B} & \textbf{Llama3-8B} & \textbf{Gemma3-12B} & \textbf{DeepSeek-R1-7B} & \textbf{Total} & \textbf{Unique} \\
\midrule[0.8pt]
Anti-VM/Sandbox & 134 & 82 & 168 & 61 & 445 & 10 \\
Evasion & 170 & 89 & 70 & 71 & 400 & 8 \\
Persistence & 18 & 13 & 13 & 2 & 46 & 3 \\
Info Stealing & 29 & 29 & 33 & 20 & 111 & 6 \\
Process Manipulation & 153 & 138 & 279 & 162 & 732 & 8 \\
System Queries & 330 & 137 & 154 & 113 & 734 & 6 \\
Network & 18 & 15 & 16 & 5 & 54 & 3 \\
Other & 303 & 177 & 219 & 154 & 853 & 38 \\
\midrule[0.5pt]
\textbf{Total} & 1155 & 680 & 952 & 588 & 3375 & 82 \\
\bottomrule[1.5pt]
\end{tabular}
\begin{tablenotes}
\item
\textbf{Count by Model}: Number of successful bypasses triggering signatures in each category.
\item
\textbf{Total}: Sum of signature triggers across all models. \textbf{Unique}: Unique signature types in category.
\end{tablenotes}

\end{threeparttable}
\label{tab:signature-categories}
\end{table*}

\newpage
\section{Complete List of Malware Families Detected by \tool}
\label{appendix:malfam}
\setlength{\tabcolsep}{3pt}
\scriptsize
\begin{longtable}{l|rr|rrrrrrr|r|r}
\caption{Malware Family Detection - Complete List}
\label{tab:family-full} \\
\toprule[1.5pt]
 & \multicolumn{2}{c|}{\textbf{Baseline}} & \multicolumn{7}{c|}{\textbf{MalwareBazaar Sandboxes}} & \textbf{Ours} & \\
\textbf{Family} & \textbf{Joe Sandbox} & \textbf{CAPE} & \textbf{ANYRUN} & \textbf{MB-CAPE} & \textbf{Intezer} & \textbf{MB-Sigature} & \textbf{Triage} & \textbf{VMRay} & \textbf{ReverseLab} & \textbf{ABLE} & \textbf{$\Delta$}$^\dagger$ \\
\midrule[0.8pt]
\endfirsthead
\multicolumn{12}{c}{\tablename\ \thetable{}} \\
\toprule[1.5pt]
 & \multicolumn{2}{c|}{\textbf{Baseline}} & \multicolumn{7}{c|}{\textbf{MalwareBazaar Sandboxes}} & \textbf{Ours} & \\
\textbf{Family} & \textbf{Joe} & \textbf{CAPE} & \textbf{ANY} & \textbf{MB-C} & \textbf{Int} & \textbf{Sig} & \textbf{Tri} & \textbf{VM} & \textbf{RL} & \textbf{ABLE} & \textbf{$\Delta$}$^\dagger$ \\
\midrule[0.8pt]
\endhead
\midrule[0.8pt]
\multicolumn{12}{r}{\textit{Continued on the next page ...}} \\
\endfoot
\bottomrule[1.5pt]
\multicolumn{12}{l}{\footnotesize $^\dagger$$\Delta$ = ABLE minus best existing sandbox count.} \\
\multicolumn{12}{l}{\footnotesize ANY = ANY.RUN, MB-C = CAPE (MalwareBazaar), Int = Intezer, Sig = MB-Signature, Tri = Triage, VM = VMRay, RL = ReversingLabs.} \\
\endlastfoot
RedLine & 196 & 195 & 128 & 183 & 41 & 108 & 131 & 178 & 74 & 240 & +44 \\
Amadey & 140 & 140 & 86 & 1 & 134 & 107 & 77 & 115 & 13 & 198 & +58 \\
Formbook & 45 & 41 & 28 & 28 & 30 & 44 & 38 & 13 & 9 & 74 & +29 \\
StealC & 9 & 10 & 0 & 0 & 3 & 3 & 3 & 3 & 0 & 39 & +29 \\
Remcos & 10 & 10 & 3 & 7 & 5 & 10 & 9 & 3 & 5 & 23 & +13 \\
STOP & 7 & 7 & 2 & 3 & 2 & 6 & 1 & 2 & 1 & 19 & +12 \\
StormKitty & 1 & 1 & 0 & 0 & 0 & 1 & 1 & 0 & 0 & 13 & +12 \\
AsyncRAT & 1 & 1 & 0 & 0 & 0 & 0 & 0 & 1 & 1 & 13 & +12 \\
WorldWind & 0 & 1 & 0 & 0 & 0 & 0 & 0 & 0 & 0 & 13 & +12 \\
PYSA & 3 & 3 & 0 & 0 & 0 & 3 & 0 & 0 & 3 & 11 & +8 \\
SystemBC & 2 & 3 & 0 & 2 & 3 & 3 & 2 & 0 & 1 & 10 & +7 \\
Cuba & 1 & 1 & 0 & 0 & 0 & 1 & 0 & 0 & 1 & 8 & +7 \\
DLAgent05 & 0 & 4 & 0 & 0 & 0 & 0 & 0 & 0 & 0 & 8 & +4 \\
Vidar & 23 & 10 & 7 & 5 & 0 & 4 & 12 & 0 & 3 & 6 & -17 \\
Clop & 0 & 1 & 0 & 0 & 0 & 0 & 0 & 0 & 0 & 6 & +5 \\
Azorult & 3 & 3 & 0 & 1 & 1 & 4 & 3 & 0 & 2 & 5 & +1 \\
WarzoneRAT & 0 & 4 & 0 & 2 & 0 & 0 & 4 & 3 & 0 & 4 & 0 \\
CobaltStrikeStager & 0 & 1 & 0 & 0 & 0 & 0 & 0 & 0 & 0 & 2 & +1 \\
NjRAT & 1 & 1 & 1 & 1 & 0 & 1 & 1 & 0 & 0 & 2 & +1 \\
Modi & 0 & 2 & 0 & 0 & 0 & 1 & 0 & 0 & 0 & 2 & 0 \\
Raccoon & 1 & 1 & 0 & 1 & 0 & 2 & 3 & 0 & 34 & 1 & -33 \\
AgentTesla & 0 & 0 & 0 & 0 & 0 & 0 & 0 & 0 & 15 & 1 & -14 \\
Disabler & 0 & 1 & 0 & 0 & 0 & 0 & 0 & 0 & 1 & 1 & 0 \\
Lumma & 5 & 1 & 0 & 0 & 0 & 1 & 0 & 3 & 0 & 1 & -4 \\
Emotet & 1 & 1 & 1 & 1 & 1 & 0 & 1 & 0 & 1 & 1 & 0 \\
Femato & 0 & 0 & 0 & 0 & 0 & 0 & 0 & 0 & 0 & 1 & +1 \\
Gh0stRAT & 1 & 1 & 1 & 1 & 1 & 1 & 1 & 1 & 1 & 1 & 0 \\
DarksideV1 & 0 & 1 & 0 & 0 & 0 & 0 & 0 & 0 & 0 & 1 & 0 \\
RunningRAT & 1 & 1 & 0 & 0 & 0 & 1 & 1 & 0 & 0 & 1 & 0 \\
Neshta & 1 & 1 & 0 & 0 & 0 & 1 & 1 & 0 & 1 & 1 & 0 \\
MedusaLocker & 1 & 1 & 1 & 1 & 1 & 0 & 0 & 0 & 0 & 1 & 0 \\
Injectorx & 0 & 0 & 0 & 0 & 0 & 0 & 0 & 0 & 0 & 1 & +1 \\
Jalapeno & 0 & 0 & 0 & 0 & 0 & 0 & 0 & 0 & 0 & 1 & +1 \\
Conti & 0 & 1 & 0 & 1 & 1 & 1 & 0 & 0 & 1 & 1 & 0 \\
Locked & 0 & 1 & 0 & 0 & 0 & 0 & 0 & 0 & 0 & 1 & 0 \\
BitRAT & 1 & 1 & 0 & 1 & 1 & 1 & 1 & 0 & 0 & 1 & 0 \\
SmokeLoader & 1 & 0 & 0 & 0 & 0 & 1 & 1 & 1 & 38 & 0 & -38 \\
Private & 0 & 0 & 0 & 0 & 0 & 0 & 0 & 0 & 49 & 0 & -49 \\
Oski & 3 & 0 & 0 & 0 & 2 & 3 & 3 & 0 & 0 & 0 & -3 \\
Tnega & 0 & 0 & 0 & 0 & 0 & 0 & 0 & 0 & 1 & 0 & -1 \\
Leonem & 0 & 0 & 0 & 0 & 0 & 0 & 0 & 0 & 9 & 0 & -9 \\
BumbleBee & 0 & 0 & 0 & 2 & 0 & 0 & 0 & 0 & 0 & 0 & -2 \\
Wacatac & 0 & 0 & 0 & 0 & 0 & 0 & 0 & 0 & 3 & 0 & -3 \\
MintZard & 0 & 0 & 0 & 0 & 0 & 0 & 0 & 0 & 1 & 0 & -1 \\
Pwsx & 0 & 0 & 0 & 0 & 0 & 0 & 0 & 0 & 1 & 0 & -1 \\
Zenpak & 0 & 0 & 0 & 0 & 0 & 0 & 0 & 0 & 3 & 0 & -3 \\
Swotter & 0 & 0 & 0 & 0 & 0 & 0 & 0 & 0 & 2 & 0 & -2 \\
GenSteal & 0 & 0 & 0 & 0 & 0 & 0 & 0 & 0 & 1 & 0 & -1 \\
SnakeLogger & 0 & 0 & 0 & 0 & 0 & 0 & 0 & 0 & 3 & 0 & -3 \\
Kelihos & 0 & 0 & 0 & 0 & 0 & 1 & 0 & 0 & 0 & 0 & -1 \\
Noon & 0 & 0 & 0 & 0 & 0 & 0 & 0 & 0 & 4 & 0 & -4 \\
Mustang Panda & 0 & 0 & 0 & 0 & 1 & 0 & 0 & 0 & 0 & 0 & -1 \\
Stealerium & 0 & 0 & 0 & 0 & 1 & 0 & 0 & 0 & 0 & 0 & -1 \\
Seraph & 0 & 0 & 0 & 0 & 0 & 0 & 0 & 0 & 2 & 0 & -2 \\
AveMaria & 4 & 0 & 3 & 0 & 3 & 4 & 0 & 0 & 2 & 0 & -4 \\
HydraCrypt & 0 & 0 & 0 & 0 & 0 & 0 & 0 & 0 & 1 & 0 & -1 \\
XMRig & 1 & 0 & 0 & 0 & 0 & 0 & 1 & 0 & 0 & 0 & -1 \\
Aurora & 0 & 0 & 0 & 0 & 1 & 0 & 1 & 0 & 0 & 0 & -1 \\
Arkei & 0 & 0 & 2 & 0 & 0 & 6 & 0 & 0 & 0 & 0 & -6 \\
Kronos & 0 & 1 & 0 & 1 & 0 & 1 & 1 & 0 & 0 & 0 & -1 \\
Konus & 0 & 0 & 0 & 0 & 0 & 0 & 0 & 0 & 1 & 0 & -1 \\
Tinba & 0 & 0 & 0 & 0 & 0 & 0 & 0 & 0 & 1 & 0 & -1 \\
Rhadamanthys & 1 & 0 & 0 & 0 & 0 & 0 & 1 & 0 & 0 & 0 & -1 \\
DarkComet & 1 & 0 & 0 & 0 & 0 & 1 & 1 & 0 & 0 & 0 & -1 \\
BlackNET & 1 & 1 & 1 & 1 & 1 & 0 & 0 & 0 & 1 & 0 & -1 \\
BlackWorm & 0 & 0 & 0 & 0 & 0 & 0 & 0 & 1 & 0 & 0 & -1 \\
Taskun & 0 & 0 & 0 & 0 & 0 & 0 & 0 & 0 & 3 & 0 & -3 \\
LokiBot & 0 & 0 & 0 & 0 & 0 & 0 & 0 & 0 & 2 & 0 & -2 \\
Babuk & 3 & 1 & 0 & 0 & 0 & 1 & 1 & 0 & 0 & 0 & -3 \\
BotX & 0 & 0 & 0 & 0 & 0 & 0 & 0 & 0 & 1 & 0 & -1 \\
Androm & 0 & 0 & 0 & 0 & 0 & 0 & 0 & 0 & 1 & 0 & -1 \\
XWorm & 0 & 0 & 0 & 0 & 0 & 0 & 1 & 0 & 0 & 0 & -1 \\
Delf & 0 & 0 & 0 & 0 & 0 & 0 & 0 & 0 & 1 & 0 & -1 \\
CryptOne & 0 & 0 & 0 & 0 & 0 & 0 & 0 & 1 & 0 & 0 & -1 \\
Mokes & 0 & 0 & 0 & 0 & 0 & 0 & 0 & 0 & 2 & 0 & -2 \\
Eternity & 0 & 0 & 0 & 0 & 0 & 0 & 1 & 0 & 0 & 0 & -1 \\
OutBreak & 0 & 0 & 0 & 0 & 0 & 0 & 0 & 0 & 1 & 0 & -1 \\
Locky & 0 & 0 & 0 & 0 & 0 & 0 & 0 & 0 & 1 & 0 & -1 \\
KeyBase & 0 & 0 & 0 & 0 & 1 & 0 & 0 & 0 & 0 & 0 & -1 \\
Spora & 0 & 0 & 0 & 0 & 0 & 0 & 0 & 0 & 1 & 0 & -1 \\
TrickBot & 0 & 0 & 0 & 0 & 0 & 1 & 0 & 0 & 0 & 0 & -1 \\
Heracles & 0 & 0 & 0 & 0 & 0 & 0 & 0 & 0 & 1 & 0 & -1 \\
zgRAT & 2 & 0 & 0 & 3 & 0 & 1 & 0 & 0 & 0 & 0 & -3 \\
Ficker & 1 & 0 & 0 & 0 & 0 & 0 & 0 & 0 & 0 & 0 & -1 \\
ClipBanker & 0 & 0 & 0 & 0 & 0 & 0 & 0 & 0 & 1 & 0 & -1 \\
Bulz & 0 & 0 & 0 & 0 & 0 & 0 & 0 & 0 & 1 & 0 & -1 \\
Multiverze & 0 & 0 & 0 & 0 & 0 & 0 & 0 & 0 & 1 & 0 & -1 \\
Fabookie & 1 & 0 & 0 & 0 & 0 & 0 & 0 & 0 & 0 & 0 & -1 \\
ManusCrypt & 1 & 0 & 0 & 0 & 0 & 0 & 0 & 0 & 0 & 0 & -1 \\
LaplasClipper & 0 & 0 & 0 & 0 & 0 & 1 & 0 & 0 & 0 & 0 & -1 \\
Zilla & 0 & 0 & 0 & 0 & 0 & 0 & 0 & 0 & 1 & 0 & -1 \\
MassLogger & 0 & 0 & 0 & 0 & 3 & 0 & 0 & 0 & 0 & 0 & -3 \\
Johnnie & 0 & 0 & 0 & 0 & 0 & 0 & 0 & 0 & 1 & 0 & -1 \\
Znyonm & 0 & 0 & 0 & 0 & 0 & 0 & 0 & 0 & 1 & 0 & -1 \\
DarkSide & 1 & 0 & 0 & 0 & 0 & 1 & 1 & 0 & 1 & 0 & -1 \\
Renamer & 0 & 1 & 0 & 1 & 0 & 0 & 0 & 0 & 0 & 0 & -1 \\
Grenam & 0 & 0 & 0 & 0 & 0 & 0 & 0 & 0 & 1 & 0 & -1 \\
Phonzy & 0 & 0 & 0 & 0 & 0 & 0 & 0 & 0 & 1 & 0 & -1 \\
Casdet & 0 & 0 & 0 & 0 & 0 & 0 & 0 & 0 & 1 & 0 & -1 \\
Pony & 2 & 2 & 2 & 1 & 2 & 2 & 2 & 2 & 2 & 0 & -2 \\
Bladabhindi & 0 & 0 & 0 & 0 & 0 & 0 & 0 & 0 & 1 & 0 & -1 \\
Farfli & 0 & 0 & 0 & 0 & 1 & 0 & 0 & 0 & 0 & 0 & -1 \\
Gold Dragon & 0 & 0 & 0 & 0 & 0 & 0 & 0 & 1 & 0 & 0 & -1 \\
ZxShell & 0 & 0 & 0 & 0 & 0 & 0 & 0 & 1 & 0 & 0 & -1 \\
MarsStealer & 1 & 0 & 0 & 0 & 0 & 1 & 2 & 0 & 2 & 0 & -2 \\
Glupteba & 0 & 0 & 0 & 0 & 0 & 0 & 0 & 0 & 1 & 0 & -1 \\
Tepfer & 0 & 0 & 0 & 0 & 0 & 0 & 0 & 0 & 1 & 0 & -1 \\
Ymacco & 0 & 0 & 0 & 0 & 0 & 0 & 0 & 0 & 1 & 0 & -1 \\
RedLineSteal & 0 & 0 & 0 & 0 & 0 & 0 & 0 & 0 & 1 & 0 & -1 \\
Rescoms & 0 & 0 & 0 & 0 & 0 & 0 & 0 & 0 & 1 & 0 & -1 \\
NanoBot & 0 & 0 & 0 & 0 & 0 & 0 & 0 & 0 & 1 & 0 & -1 \\
TeamBot & 0 & 0 & 0 & 0 & 0 & 1 & 0 & 0 & 0 & 0 & -1 \\
CobaltStrike & 1 & 0 & 0 & 0 & 0 & 1 & 0 & 0 & 0 & 0 & -1 \\
Metasploit & 1 & 0 & 0 & 0 & 0 & 0 & 1 & 0 & 0 & 0 & -1 \\
ContiCrypt & 0 & 0 & 0 & 0 & 0 & 0 & 0 & 0 & 1 & 0 & -1 \\
DBat & 2 & 0 & 0 & 0 & 0 & 0 & 0 & 0 & 0 & 0 & -2 \\
modi & 0 & 0 & 0 & 0 & 0 & 0 & 1 & 0 & 0 & 0 & -1 \\
NanoCore & 0 & 0 & 0 & 0 & 0 & 0 & 0 & 0 & 1 & 0 & -1 \\
Guildma & 0 & 0 & 0 & 0 & 0 & 0 & 0 & 0 & 1 & 0 & -1 \\
BlueSky & 1 & 0 & 0 & 0 & 0 & 0 & 0 & 0 & 0 & 0 & -1 \\
Racealer & 0 & 0 & 0 & 0 & 0 & 0 & 0 & 0 & 1 & 0 & -1 \\
ParalaxRat & 0 & 0 & 0 & 0 & 0 & 0 & 0 & 0 & 1 & 0 & -1 \\
DCRat & 1 & 0 & 0 & 0 & 0 & 1 & 0 & 0 & 0 & 0 & -1 \\
Deyma & 0 & 0 & 0 & 0 & 0 & 0 & 0 & 0 & 1 & 0 & -1 \\
\midrule[0.8pt]
\textbf{Total} & \textbf{484} & \textbf{457} & \textbf{266} & \textbf{248} & \textbf{240} & \textbf{332} & \textbf{309} & \textbf{329} & \textbf{331} & \textbf{712} & \textbf{+228} \\
\end{longtable}
\normalsize
\setlength{\tabcolsep}{6pt}

\newpage

\end{document}